\documentclass[iop]{emulateapj}

\usepackage{amsmath}
\usepackage{apjfonts}
\usepackage{amssymb}
\usepackage{natbib}
\usepackage{multirow}
\usepackage{mathrsfs}
\usepackage{color}
\usepackage{hyperref}
\defcitealias{Faber:2005p4315}{FRW}
\defcitealias{Press:1977p3083}{PT}

\shorttitle{Ejection and Disruption of Giant Planets}
\shortauthors{Guillochon, Ramirez-Ruiz \& Lin}

\begin{document}
\title{Consequences of the Ejection and Disruption of Giant Planets}
\author{James Guillochon\altaffilmark{1,2}, Enrico Ramirez-Ruiz\altaffilmark{1}, and Douglas Lin\altaffilmark{1}}
\altaffiltext{1}{Theoretical Astrophysics Santa Cruz (TASC), Department of Astronomy and Astrophysics, University of California, Santa Cruz, CA 95064, USA; \email{jfg@ucolick.org}}
\altaffiltext{2}{NASA Earth and Space Science Fellow}
\begin{abstract}
The discovery of Jupiter-mass planets in close orbits about their parent stars has challenged models of planet formation. Recent observations have shown that a number of these planets have highly inclined, sometimes retrograde orbits about their parent stars, prompting much speculation as to their origin. It is known that migration alone cannot account for the observed population of these misaligned hot Jupiters, which suggests that dynamical processes after the gas disc dissipates play a substantial role in yielding the observed inclination and eccentricity distributions. One particularly promising candidate is planet-planet scattering, which is not very well understood in the non-linear regime of tides. Through three-dimensional hydrodynamical simulations of multi-orbit encounters, we show that planets that are scattered into an orbit about their parent stars with closest approach distance being less than approximately three times the tidal radius are either destroyed or completely ejected from the system. We find that as few as 5 and as many as 18 of the currently known hot Jupiters have a maximum initial apastron for scattering that lies well within the ice line, implying that these planets must have migrated either before or after the scattering event that brought them to their current positions. If stellar tides are unimportant $(Q_\ast \gtrsim 10^7)$, disk migration is required to explain the existence of the hot Jupiters present in these systems. Additionally, we find that the disruption and/or ejection of Jupiter-mass planets deposits a Sun's worth of angular momentum onto the host star. For systems in which planet-planet scattering is common, we predict that planetary hosts have up to a 35\% chance of possessing an obliquity relative to the invariable plane of greater than 90 degrees.
\end{abstract}
\keywords{Planet-star Interactions --- Hydrodynamics --- Gravitation --- Chaos --- Stars: rotation --- Ultraviolet: planetary systems}
\section{Introduction}
The search for planets about other stars has led to the discovery of dozens of planets with unusual properties. As both radial velocity and transit surveys are biased towards planets that are both massive and close to their parent stars, the region of parameter space corresponding to planets with a mass larger than that of Neptune and a semi-major axis $< 0.1$ au is particularly well-explored \citep{Ida:2004p4433, Shen:2008p4684, Zakamska:2010p5800}. This has led to the discovery of many giant planets known colloquially as hot Jupiters and Neptunes, which are thought to have formed far away from their parent stars but were then later transplanted to their observed positions by currently undetermined means. Many of these exoplanets come so close to their parent stars that they toe the line between destruction and survival, with some observed exoplanets in danger of being destroyed on a relatively short timescale \citep{Li:2010p4958}. Additionally, the inclination distribution of the hot Jupiters seems to demonstrate significant misalignment between the planet's orbit and the stellar spin axis \citep{Triaud:2010p4772, Schlaufman:2010p4689}, a surprising result that may require a dynamical process that acts after the protoplanetary disk dissipates.

There are three primary physical processes that can deposit a planet on an orbit that is very close to its parent star: disk migration, the Kozai mechanism, and planet-planet scattering. Disk migration can yield hot Jupiters, but as the star collapses from the same cloud as the protoplanetary disk that encircles it, it is difficult to explain the observed orbit misalignments using this mechanism alone \citep[though see][for discussions  regarding star-disk interactions]{Foucart:2010p4925,Watson:2010p4924}. The Kozai mechanism \citep{Kozai:1962p5046} can lead to the large eccentricities required to produce close-in planets, but it can only operate in systems in which a massive planet or secondary star are present, and the mechanism may be mitigated by general relativistic effects that become important before tidal dissipation is large enough to circularize the orbit \citep{Takeda:2005p4492, Fabrycky:2007p5801}. Planet-planet scattering can produce both the observed semi-major axis and inclination distributions, and can deposit planets close enough such that tides can circularize the orbits in a time that is less than the system age. Additionally, the object that acts as a scatterer can have approximately the same mass as the scattered object itself and still yield a hot Jupiter in a significant fraction of systems \citep{Ford:2008p4328}, negating the need for a non-planetary companion in the system.

Previous hydrodynamical work has only focused on the planet's first close fly-by \citep[hereafter FRW]{Faber:2005p4315}, and does not investigate how prolonged tidal forcing over many orbits affects a planet's chances for survival. In this paper we have performed hydrodynamical simulations of multiple passages of a Jupiter-like planet by a Sun-like star, bridging the gap between numerical and analytical work that have focused on extremely close and extremely grazing encounters respectively. We find that scattering planets into star-grazing orbits is more destructive than previously thought, with Jupiter-like planets being destroyed or ejected at distances no smaller than 2.7 times the tidal radius \(r_{\rm t} \equiv R_{\rm P} (M_\ast/M_{\rm P})^{1/3}\). As some exoplanets are currently observed to have semi-major axes less than twice this critical value, their initial eccentricities may be required to have been substantially smaller than unity if planet-planet scattering is the mechanism responsible for bringing them so close to their host stars. This strongly suggests that planet-planet scattering alone cannot explain the complete observed population of close-in Jupiter-like exoplanets, and that the process must operate along with one of either the Kozai mechanism, disk migration, or both. These three processes likely act in concert to produce the observed population of hot gas giants, with the relative importance of each process being a function of the system's initial conditions.

If planet-planet scattering is common enough to explain the existence of hot Jupiters, we predict that there should be two signatures of disruption that are readily detectable with today's instruments. Firstly, we find that the parent star can have its spin significantly altered by the accretion of material removed from the planet as a result of the disruption, producing a star that can be significantly misaligned relative to any remaining planets. Secondly, we find that most planet disruption events lead to the planet's ejection from the host system prior to the planet being completely destroyed, and that this ejected planet can remain almost as bright as its host star for centuries.

In this paper we focus on the results of numerical hydrodynamical simulations that have been used to attempt to ascertain the true radii of destruction and ejection for Jupiter-like exoplanets, and the consequences of these planet-removing processes on their stellar hosts. In Section \ref{sec:modeling} we review the history of the analytical and numerical work done to characterize the orbital evolution of a planet that comes within a few tidal radii of its host star, and then we detail our particular numerical approach to modeling tidal disruption. We report the results of our simulations in Section \ref{sec:simres}. In Section \ref{sec:disc} we discuss the implications of our results, with special attention paid to the viability of various mechanisms for producing hot Jupiters, and the observational signatures of planetary disruption and ejection. We summarize the shortcomings of our models and the possible fates of a Jupiter-like exoplanet in Section \ref{sec:conclusions}. Appendix \ref{sec:modgrav} is provided to detail our algorithm used to simulate multiple orbits and for presenting tests of the algorithm's conservative properties.

\section{Modeling Planetary Disruption}\label{sec:modeling}
\begin{figure*}[t]
\centering\includegraphics[width=\linewidth,clip=true]{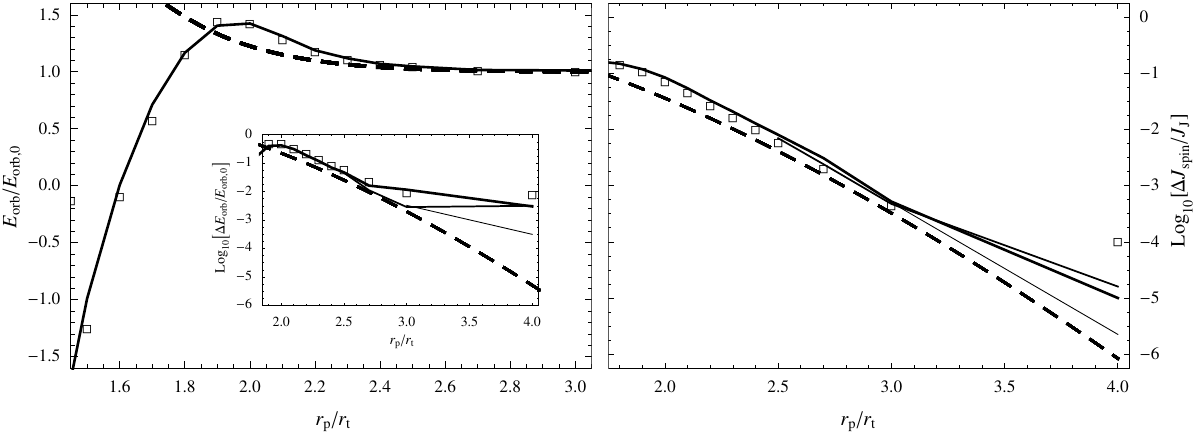}
\caption{Final orbital energy $E_{\rm orb}$ scaled to the initial orbital energy $E_{\rm orb, 0}$ (left panel), change in orbital energy $\Delta E_{\rm orb} \equiv E_{\rm orb, 0} - E_{\rm orb}$ scaled by $E_{\rm orb, 0}$ (left sub-panel), and spin angular momentum $J_{\rm spin}$ scaled by the characteristic angular momentum of Jupiter \(J_{\rm J}^2 \equiv G M_{\rm J}^3 R_{\rm J}\) (right panel) as functions of periastron distance $r_{\rm p}$ after a single near-parabolic encounter between a $M_{\rm P} = M_{\rm J}$ Jupiter-like planet and a $M_\ast = 10^3 M_{\rm J}$ star. The solid lines show the results from this work, with decreasing thickness corresponding to increasing maximum refinement, square markers show the results of \citetalias{Faber:2005p4315}, and the dashed lines shows the prediction of \citetalias{Press:1977p3083} using \citetalias{Faber:2005p4315}'s analytical $n = 1$ solutions for $T_{l}$ and $S_{l}$. Note that our results and that of \citetalias{Faber:2005p4315} are consistent for $r_{\rm p} \le 2.5 r_{\rm t}$. At the highest resolution, our results are consistent with that of \citetalias{Press:1977p3083} until $r_{\rm p} \ge 4 r_{\rm t}$.}
\label{fig:frwcomp}
\end{figure*}

\begin{figure*}[t]
\centering\includegraphics[width=\linewidth,clip=true]{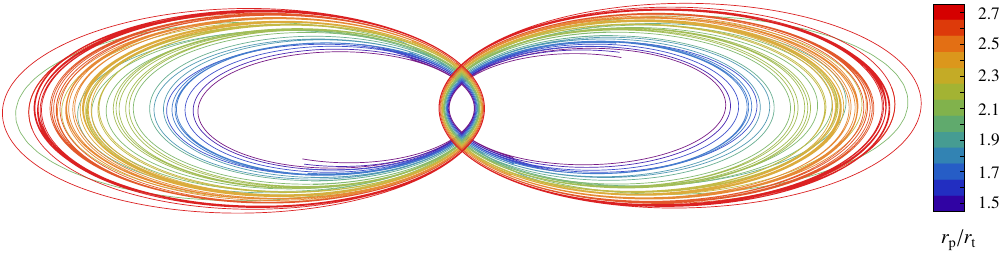}
\caption{The double rainbow curves show orbital trajectories for both the planet (left) and star (right) during the encounter for different values of the periastron distance $r_{\rm p}$. The star's trajectory is magnified by a factor of $M_{\ast}/M_{\rm P} = 10^3$ to make its motion apparent. While the orbits do precess slightly over the course of the simulations, this plot shows the orbits with the precession removed. Note that the $r_{p} = 21$ run (light green) experiences a particularly ejective encounter on its 2nd periastron passage.}
\label{fig:trajectories}
\end{figure*}

\subsection{Previous Disruption Models}\label{sec:prevmodels}
Tidal encounters between a point mass and an extended, initially spherical object have been described with progressively more detailed analytical models and simulations. The first analytical models of tidal dissipation were laid out in \cite{Press:1977p3083} (hereafter PT), which assumes that the tidally excited body retains its spherical shape and that the motions induced by the encounter can be described with spherical harmonics. This model works quite well for encounters where tides are weak, i.e. more than a few tidal radii away from the perturbing body, where the assumption of sphericity is valid. If mode-mode coupling is weak, the initially excited modes have no mechanism to share energy with each other during the encounter itself, and the \citetalias{Press:1977p3083} formalism can accurately predict the amount of energy injected into the extended object.

In the \citetalias{Press:1977p3083} formalism, the amount of energy and angular momentum injected into an object is derived by taking the convolution of the object's modes of oscillation and the frequency decomposition of the tidal field, which drops the time-dependence from the governing equations. However, if the incoming object is already oscillating, the phase of these oscillations relative to the time of periastron is important, which by construction cannot be resolved by the time-independent \citetalias{Press:1977p3083} formalism. The absence of time-dependence also requires that the base-state is cylindrically symmetric relative to the line connecting the extended object to the perturber, which can only be accomplished for an initially non-rotating object by using a spherical geometry. As the amplitude of the mode oscillations approaches the size of the object itself, the body must become highly non-spherical and therefore the \citetalias{Press:1977p3083} formalism must break down.

These two shortcomings lead to the development of the ``affine'' model for tidal encounters \citep{Carter:1983p12, Carter:1985p238}, which allows the initially spherical object to deform into a triaxial ellipsoid during and after the encounter. One axis of the ellipsoid is fixed to be perpendicular to the orbital plane, with the other two axes lying within the plane at a right angle to each other and with arbitrary orientation angle relative to the first axis. This feature allows for the object to be followed dynamically, which means that the incoming state of the object's oscillations can affect the object's response to future encounters.

Because the excitation of normal modes in a dynamically inert object can only yield an increase in the object's energy and angular momentum budgets, an initially spherical object would always find itself in a more-tightly bound orbit after the encounter. An important facet of the problem that the affine model can investigate is how the fundamental modes excited in previous encounters are de-excited in future encounters, which can lead to a positive change in the orbital energy. The inclusion of de-excitation of modes adds a chaotic component to the problem, with the orbital evolution behaving as a ``random-walk'' process \citep{Kochanek:1992p4680, Mardling:1995p4495}.

While mass loss has been included in a nested-ellipsoid version of the affine model presented by \cite{Ivanov:2001p4490}, asymmetrical mass loss is not treated as the models assume mirror symmetry. Asymmetric mass loss is expected in real disruptions as the tidal field is stronger on the side of the object facing the point mass. This is due to the steep power-law dependence of the strength of the tidal field $(\propto r^{-3})$ and the non-linear evolution of the tide raised on the object, with both the velocity and amount of mass lost being larger on the side of the object closest to the point mass. Numerical simulations by \citetalias{Faber:2005p4315} show that this asymmetric mass loss leads to a substantial deviation from the nested-ellipsoid treatment, especially when the closest approach distance is $\lesssim 2 r_{\rm t}$. Within this distance, the asymmetric removal of mass can lead to a positive increase in the orbital energy, which can cause the object to be completely ejected from the system if $E_{\rm orb} \equiv -G M_\ast M_{\rm P} / 2 a \sim E_{\rm obj}$. 

The disruption of polytropes in highly eccentric encounters has been investigated numerically by both Lagrangian \citep{Nolthenius:1982p5484,Bicknell:1983p604,Evans:1989p147,Kobayashi:2004p152,Faber:2005p4315,Rosswog:2008p3059,Rosswog:2009p3553,RamirezRuiz:2009p3071,Lodato:2009p3051} and Eulerian methods \citep{Khokhlov:1993p5432,Khokhlov:1993p5430,Frolov:1994p2,Diener:1997p457,Guillochon:2009p3441}, with the principle focus being on stars or compact objects that are disrupted by point-like gravitational sources. However, most hydrodynamic simulations that have focused on the long-term survival of these systems only consider when the two objects have a mass ratio close to unity \citep{Lee:2010p4412, LorenAguilar:2009p4672}, or for non-disruptive encounters \citep{Rathore:2005p4797}. Recently, \cite{Antonini:2010p4875} performed low-resolution multiple-passage simulations in the context of the galactic center, allowing the exploration of a large parameter space at the expense of accuracy.

To summarize, most analytical models of tides in an astrophysical context focus on objects which do not lose any mass at their closest approach, and conversely most numerical work has focused on encounters where the object is completely destroyed. The intermediate regime, where objects lose some mass but are not completely destroyed in their first passage, is largely uncharacterized. If the planet survives the initial encounter, some of the mass that is removed from the planet can become bound to the planet again as it recedes from periastron to a region with a weaker tidal field. The return and subsequent re-accretion of this material is not treated at all in analytical models. As we will describe in Section \ref{sec:multpass}, the structure of the planet can be significantly altered by the re-accretion of the planetary envelope, and the inclusion of this re-accretion into tidal disruption theory is necessary to determine if a planet will ultimately survive.

\subsection{Our Approach}\label{sec:ourapproach}
While Lagrangian codes are well-suited for treating problems where a hydrostatic object moves rapidly with respect to a fixed reference frame, their relatively poor scalability makes it difficult to follow the evolution of an object for many dynamical timescales with sufficient spatial resolution. On the other hand, maintaining near-hydrostatic balance in rapidly advecting frames using Eulerian methods can also be quite challenging. \cite{Robertson:2009p3764} show that performing Eulerian simulations in a boosted frame with unresolved pressure gradients leads to a spurious viscosity term that tends to damp out instabilities. This has been colloquially referred to as the non-Galilean invariance (GI) of the Riemann problem \citep{Springel:2010p1909,Tasker:2008p4296}. While \citeauthor{Robertson:2009p3764} showed that GI issues can be avoided with increased resolution, the requisite spatial resolution can be impractical for three-dimensional simulations that are evolved over thousands of dynamical timescales, or for when the bulk velocities are many times larger than the internal sound speeds within the simulation. The problem is compounded in the outermost layers of Jupiter-like planets, which exhibit particularly steep pressure gradients.

To side-step the GI issues, our simulations are performed in the rest-frame of the planet, which limits the typical fluid velocities to the planet's sound speed for tidal disruption calculations \citep{Guillochon:2009p3441}. We model the planet as an $n = 1$, $\Gamma = 2$ polytrope, with the fluid being described by a polytropic equation of state $(P \propto \rho^\gamma)$, where the adiabatic index $\gamma = \Gamma$. This gives a reasonable approximation to the interior structure of Jupiter-like planets \citep{Hubbard:1984p5843}, as long as the core is not a significant fraction of the planet's mass. Our simulations are constructed within the framework of FLASH \citep{Fryxell:2000p440}, an adaptive-mesh, grid-based hydrodynamics code. We treat the star as a point-mass because the distortion of the star itself contributes negligibly to the planet's orbital evolution in the case where $M_{\ast} \gg M_{\rm P}$ \citep{Matsumura:2008p4494}. The relative positions of the star and the planet are explicitly tracked over the course of the simulation using two virtual particles $\mathbf{x}_{\rm S}$ (star) and $\mathbf{x}_{\rm P}$ (planet), which are evolved alongside the hydro calculation using a Burlisch-Stoer integrator \citep{Press:1986p1001} with maximum error constrained to be less than $10^{-12}$. The planet's self-gravity is calculated using a multipole $O(l_{\max}^{2} N)$ expansion of the fluid, where we take $l_{\max} = 12$.

The flux-calculation step in any hydro code depends on the time-dependent force applied to the fluid over the duration of the step. Because the gravitational field is applied as a source term in most multi-dimensional hydro codes that include self-gravity, conservation of energy is not explicitly achievable when self-gravity is included, as the potential at timestep $m + 1$ is unknown and must be estimated. In the case of self-bound objects in hydrostatic equilibrium, small perturbations can lead to a systematic drift in the total energy of the system. For hydrostatic objects, we find that an increase in spatial resolution reduces this drift by an amount that approximately scales with the number of voxels used to resolve a given region. This is because the relative forces acting on neighboring cells decrease as the voxels are more closely packed together, and because the sharp density gradients present in the 1D hydrostatic profile are better resolved.

The potential at $m + 1$ must be estimated because the value of $\phi^{m+1}$ depends on $\rho^{m+1}$, which is not known until the hydro step has been completed. This extrapolation is one of the main sources of error in the code because the obtained acceleration is only an estimate based on the previous evolution of $\phi$. And because the extrapolation is only first-order accurate in time, smaller time-steps do not improve the accuracy of the results \citep{Springel:2010p1909}. The error can be somewhat reduced by using a higher-order extrapolation that includes $N$ previous time-steps, but our tests show that including the $m - 2$ time-step only leads to a few percent reduction in error relative to the first-order approximation. Additionally, it imposes additional memory and disk overhead to save the full potential field from the previous $N$ time-steps.

For cases where the object is nearly in hydrostatic equilibrium the gravitational field does not rapidly change with time. However, a tidal disruption can lead to variations in the gravitational field of order unity over a fraction of the planet's dynamical timescale. As a result, the default method used by FLASH to apply the gravitational field can lead to substantial changes in the total energy. Motivated by this, we implement a novel method where the potential contribution from the fluid and from tidal forces are separated, negating much of the error associated with the extrapolation of the potential. Details of this algorithm are provided in Appendix \ref{sec:modgrav}.

\section{Simulation Results}\label{sec:simres}
\subsection{Single Passage Encounters}
\begin{figure}[t]
\centering\includegraphics[width=\linewidth,clip=true]{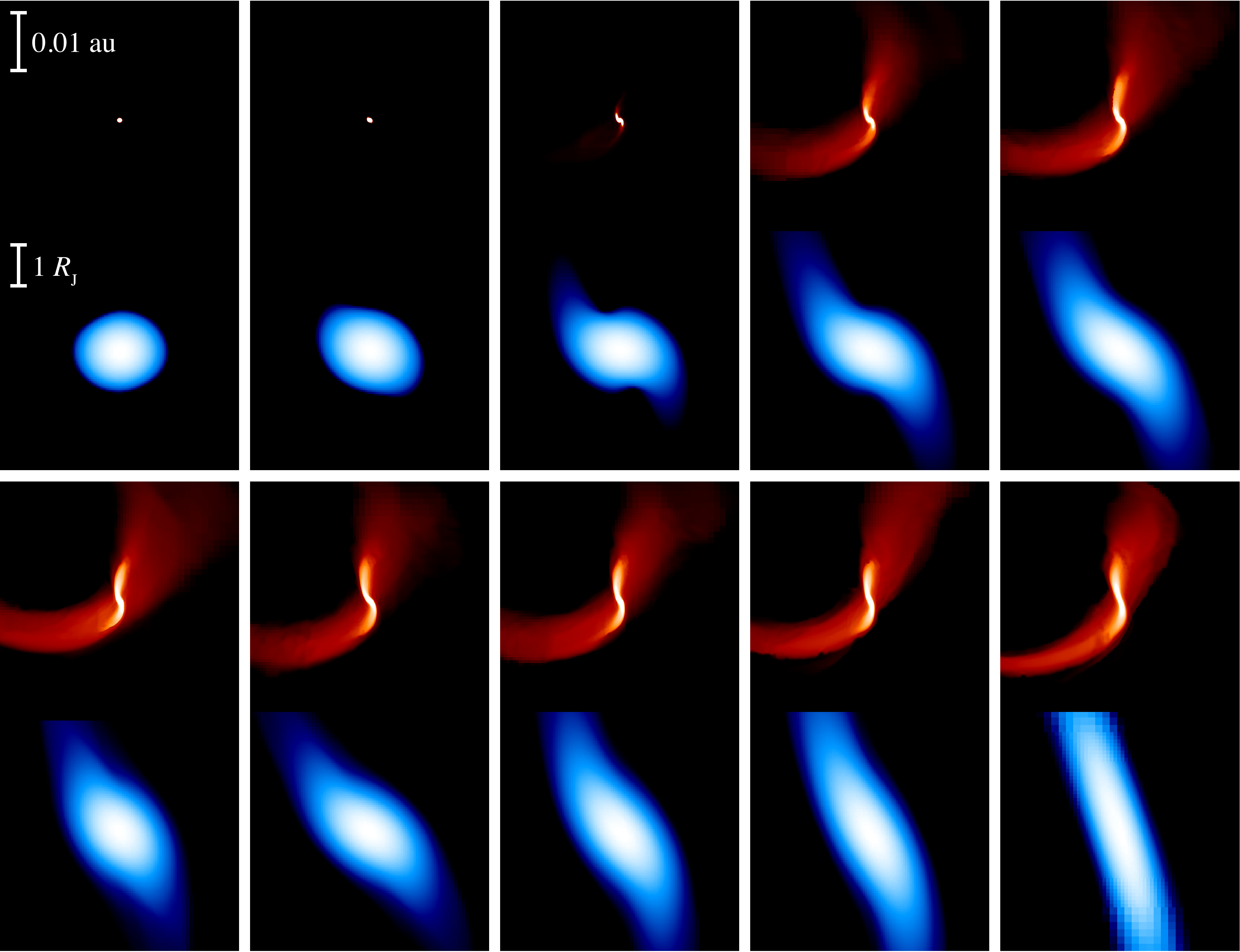}
\caption{Slices through the orbital plane shortly after each periastron passage for the simulation where $r_{\rm p} = 2.7 r_{\rm t}$. All plots show log $\rho$. The upper, red color-coded figures show a wide view of each encounter, with white corresponding to $\rho = 10^{-2}$ g cm$^{-3}$ and black corresponding to $\rho = 10^{-7}$ g cm$^{-3}$, while the lower, blue color-coded figures show a close-up view of the core, with white corresponding to the maximum density $\rho_{\max}$ and black corresponding to $\rho = 10^{-2}$ g cm$^{-3}$.}
\label{fig:multisnapshot1}
\end{figure}
\begin{figure}[t]
\centering\includegraphics[width=\linewidth,clip=true]{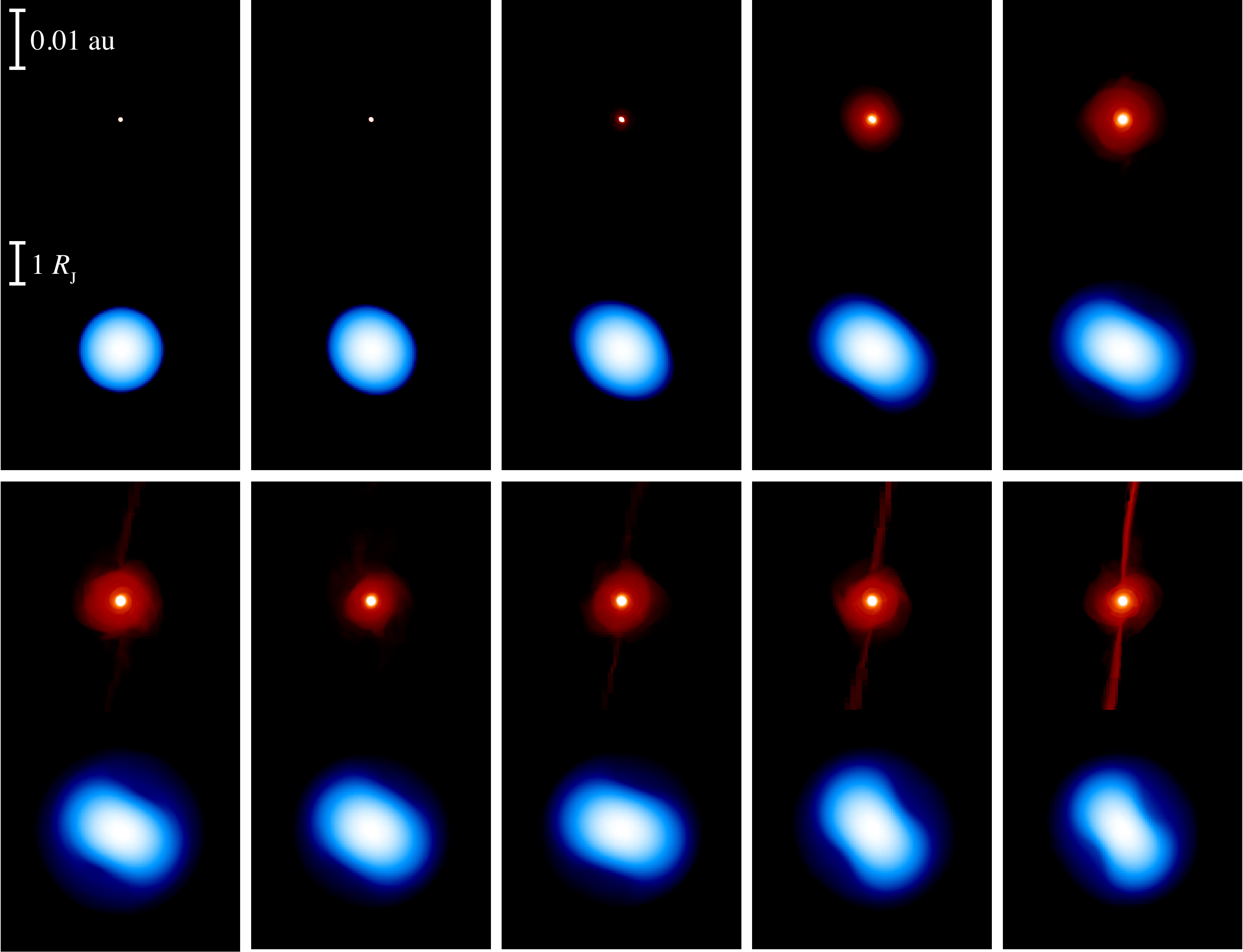}
\caption{Same as Figure \ref{fig:multisnapshot1}, but with frames 2-10 corresponding to apastron. The first frame shows the initial conditions.}
\label{fig:multisnapshot2}
\end{figure}

For comparison purposes, we first ran a suite of simulations with physical initial conditions identical to that of \citetalias{Faber:2005p4315}. The planet is assumed to have a radius $R_{\rm P} = R_{\rm J}$ and mass $M_{\rm P} = M_{\rm J}$, where $R_{\rm J}$ and $M_{\rm J}$ are the radius and mass of Jupiter. The planets are disrupted by a star with $M_\ast = 10^3 M_{\rm J}$, with the orbits of incoming planets are set to have an apastron separation $r_{\rm a} = 10^{4} R_{\rm J}$. Our lowest-resolution models have maximum spatial resolutions of $s = 0.02 R_{\rm J}$, where $s$ is the width of the smallest grid cells, corresponding to $N^{3} \simeq 10^{6}$, slightly better than the peak spatial resolution of \citetalias{Faber:2005p4315}. Our results agree very well with \citetalias{Faber:2005p4315}'s results for periastron passage distances $r_{\rm p} \leq 2.5 r_{\rm t}$ (Figure \ref{fig:frwcomp}). 

Because the amount of energy stored in the oscillations is \(\Delta E \sim E_{\rm J} (\Delta R / R_{\rm J})^2\) where $E_{\rm J}$ is the binding energy of Jupiter and $\Delta R$ is the amplitude of the mode, the simulations converge to the analytical solution only when the spatial resolution is fine enough to resolve $\Delta R$. As is evident in both \citetalias{Faber:2005p4315} and our lower-resolution runs, the measured amount of energy dissipation is larger than the analytical predictions of \citetalias{Press:1977p3083} when the oscillatory amplitude is smaller than the minimum grid scale. To test for convergent behavior, we ran higher-resolution simulations with double and quadrupole the resolution of our fiducial test for the more grazing encounters. For the $r_{\rm p} = 2.7 r_{\rm t}$ and $r_{\rm p} = 3 r_{\rm t}$ runs, the improved spatial resolution allows us to recover the analytical solution. It is also apparent that our $r_{\rm p} = 4 r_{\rm t}$ simulation is closer to convergence than the lower-resolution models, but the estimated resolution required for true convergence ($l \sim 10^{-3} R_{\rm J}, N^{3} = 10^{10}$) means that recovering the predicted results of \citetalias{Press:1977p3083} for such a grazing passage is currently beyond our computational ability.

As in \citetalias{Faber:2005p4315} and \cite{Khokhlov:1993p5432}, we also find that the change in $E_{\rm orb}$ and $J_{\rm spin}$ is slightly larger than what is predicted by the analytical models, even in the simulations that have surely converged and have minimal mass loss (i.e., $2 r_{\rm t} \le r_{\rm p} \le 2.7 r_{\rm t}$). This is almost certainly due to the dynamical tide effects neglected by the \citetalias{Press:1977p3083} model. Comparing the hydrodynamic results to that of a dynamical treatment of tides shows better agreement for this range of pericenter distances \citep{Lai:1994p3305}, but note that these dynamical models only include the $l = 2$ $f$-mode of oscillation, and thus do not account for energy transferred to higher-order $f$-modes or $p$-modes. For grazing encounters, the dynamical tide is much less important, and thus we expect that the change in orbital energy and angular momentum should converge to the \citetalias{Press:1977p3083} prediction for an inviscid planet.

\subsection{Multiple Passage Encounters}\label{sec:multpass}
\begin{figure}[t]
\centering\includegraphics[width=\linewidth,clip=true]{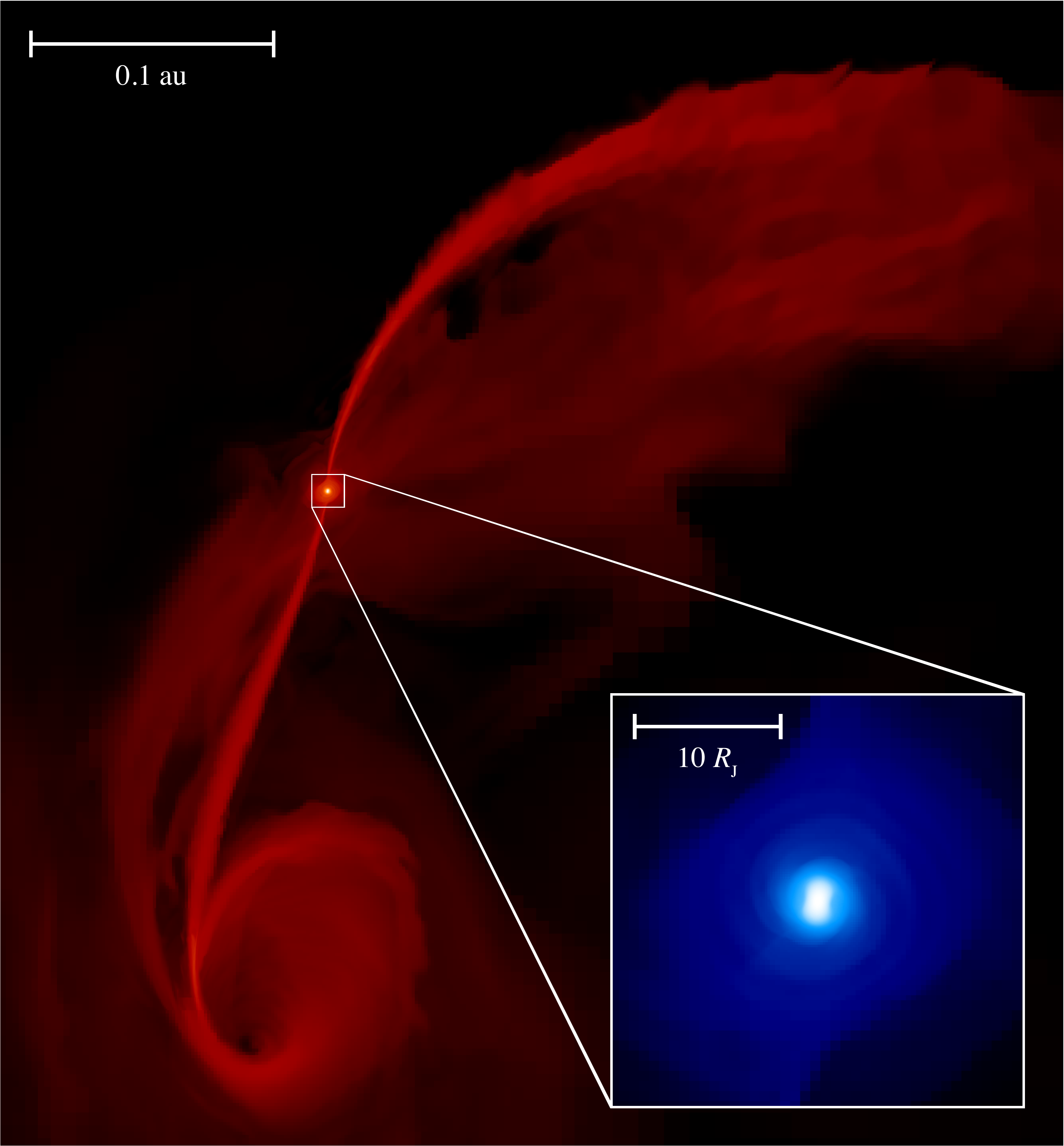}
\caption{Fallback accretion stream formed as the result of the 8$^{\rm th}$ encounter between a 1 $M_{\odot}$ star and 1 $M_{\rm J}$ planet with initial $r_{\rm p} = 2.7 r_{\rm t}$. The large, red color-coded image shows a wide-view of the disruption, while the inset blue color-coded figure shows a close-up of the surviving planetary core.}
\label{fig:tidaltails}
\end{figure}
The initial conditions of \citetalias{Faber:2005p4315} are not appropriate for investigating multiple-passage encounters as the orbital timescale is many thousands of times longer than the dynamical timescale. Thus, for our second set of simulations in which we explore multiple passages, we assume that the planets are initially on $e = 0.9$ orbits (Figure \ref{fig:trajectories}). Depending on the initial pericenter distance, the planets are destroyed in as few as one and as many as ten orbits (Figures \ref{fig:multisnapshot1} and \ref{fig:multisnapshot2}), where a planet is considered to be destroyed once it has lost more than 90\% of its original mass.

To ensure that our assumption of a less eccentric orbit is valid, we consider the effects of varying $e$ when $e \approx 1$. For nearly-parabolic orbits, the shape of the orbit in the vicinity of periastron changes very little with changing $e$, with the main effect being that the average strength of the tidal force is slightly stronger for smaller $e$ before and after pericenter. This means that the critical distance for which a planet is destroyed or ejected is very slightly larger than our setup for single encounters described in the previous section where $e \simeq 1$. To estimate the magnitude of this effect, we calculate the ratio of the distances at fixed true anomaly for two orbits with eccentricities $e_1$ and $e_2$, assuming both orbits have the same $r_{\rm p}$
\begin{equation}
    \frac{r_2}{r_1} = \frac{e_1 \left(e_2 + 1\right) r_{\rm p}}{\left(e_1 - e_2\right)r_1 + \left(e_1 + 1\right)e_2 r_{\rm p}}
\end{equation}
As the strength of the tidal force is $\propto r^{-3}$, the difference in the strength of the tidal force between the two orbits with respect to the tidal force experienced at pericenter is
\begin{align}
    \frac{F_1 - F_2}{F_{r_{\rm p}}} &= r_{\rm p}^{3}\left(r_1^{-3} - r_2^{-3}\right) & \\
    &= \frac{3\left(e_1 - e_2\right)}{e_2\left(e_2+1\right)}\frac{r_{\rm p}^3}{r_1^3}\left(\frac{r_1}{r_{\rm p}} - 1\right) + O(2) & : e_1 - e_2 \rightarrow 0.
\end{align}
This expression is maximized at $r_1 = \frac{3}{2}r_{\rm p}$, where the force difference between an $e_1 = 0.9$ and $e_2 = 1.1$ encounter evaluates to $4\%$. For all other values of $r_1$, the force differences are much smaller than this maximum. Thus, our hydrodynamical treatment of tides in a $e = 0.9$ orbit is directly applicable to all orbits with $0.9 \lesssim e \lesssim 1.1$.

After an orbit in which the planet sheds mass, some of the material that is removed from the planet's surface remains marginally bound to the planetary core. The majority of this material is then re-accreted by the planet over a few dynamical timescales. When the free-falling material encounters the surviving planetary remnant, its kinetic energy is converted to internal energy in a standing accretion shock, which results in the planet possessing a hot outer layer with temperature close to the virial temperature (Figure \ref{fig:tidaltails}). Additionally, the material striking the remnant leads to some heating of the remnant's outer mass shells. This effect is predominantly important in determining the envelope's temperature early in the accretion history before the pressure of the growing hot atmosphere becomes comparable to the ram pressure and is able to halt the flow before it can reach the core's surface \citep{Frank:2002p4864}.

\begin{figure}[tb]
\centering\includegraphics[width=\linewidth,clip=true]{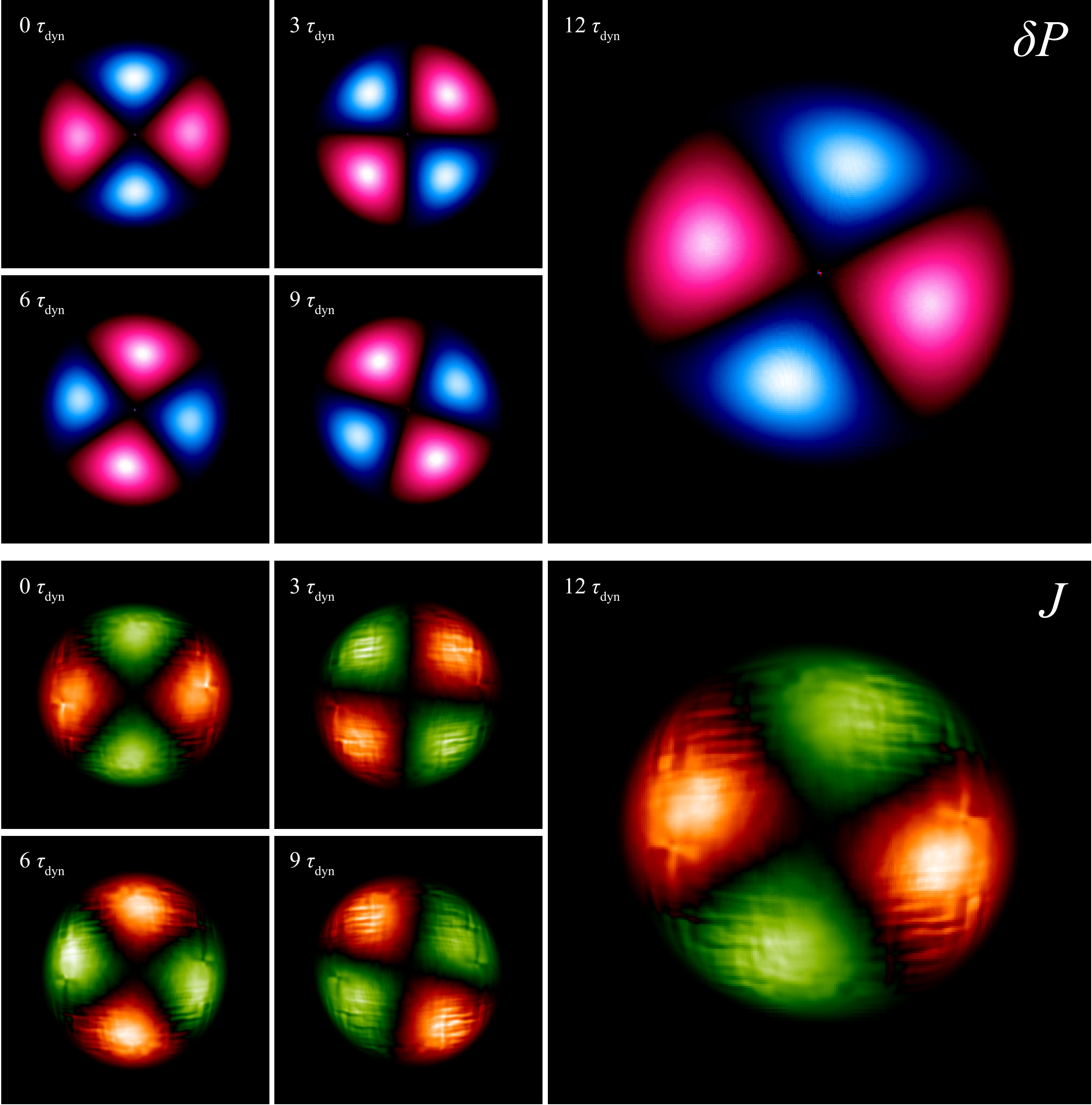}
\caption{Pressure perturbations and spin excited by a grazing encounter with $r_{\rm p} = 3 r_{\rm t}$. The upper five panels show the pressure perturbation $\delta P$ over several $\tau_{\rm dyn}$, with blue corresponding to regions of lower pressure and pink corresponding to regions of higher pressure relative to the base state. The lower five panels show the fluid's angular momentum relative to $r_{\rm c}$, with green representing clockwise rotation and orange representing counter-clockwise. Note that while the object appears to be rotating rapidly, the presence of rotating and counter-rotating regions leads to almost an exact cancellation of the total angular momentum $J$. The illusion of rapid rotation is in fact related to the pattern speed of the $l = 2$, $m = \pm2$ normal modes. Because the angular momentum carried by this mode is related to the tangential component of the expansion and contraction of the planet as it oscillates, the fluid vacillates back and forth at the mode frequency. As can be seen in the figure, the fluid with $J > 0$ possess slightly more total angular momentum than the fluid with $J < 0$. This leads to a spin frequency $\omega = \sum J/I$ that is actually very small.}
\label{fig:prespert}
\end{figure}

Because the re-accreted material is marginally bound to planet, the orbital trajectories characterizing the accretion streams have apocenter distances comparable to the planet's Hill sphere \(r_{\rm H} \equiv r_{\rm p} (M_{\rm P}/3 M_{\ast})^{1/3}\) and follow highly eccentric orbits. As the re-accreted material returns on a Keplerian trajectory \citep{Kochanek:1994p671, RamirezRuiz:2009p3071}, it carries a substantial amount of specific angular momentum. This leads to a rapid spin-up of the planetary remnant's outer layers. In encounters with little or no mass loss, the planets spin slowly post-encounter as the angular momentum carried by the normal modes is almost equally distributed between rotating and counter-rotating regions (Figure \ref{fig:prespert}). The process of re-accretion produces a planet that has a lower average density and more mass at larger radii, which makes the planet easier to destroy on subsequent passages.

As the re-accreted material has a temperature comparable to the virial temperature, radiative cooling may be able affect the atmosphere's structure, an effect that we do not account for in our simulations. However, as the planet can only thermally evolve for one orbital period, the atmosphere does not have much time to cool down before it has another strong tidal encounter with the star. While the outer layers of the planet may cool relatively rapidly and lead produce a brief transient visible in the UV (see Section \ref{sec:obsig}), the denser regions of the planet's hot atmosphere component contains much more mass and is too optically thick to cool significantly before returning to pericenter. Assuming Thomson scattering, this corresponds to a mass of the hot atmosphere component $M_{\rm atm} \sim 10^{-5} M_{\rm J}$. This means that we expect that the thermal evolution of the reaccreted material is unimportant in determining the planet's density profile interior to the most tenuous outer layers. Thus, we expect that the dynamically relevant distribution of mass within the planet should remain relatively unchanged between encounters with the star, even for values of $e$ significantly closer to 1 than what we use in our simulations.

\begin{figure}[t]
\centering\includegraphics[width=\linewidth,clip=true]{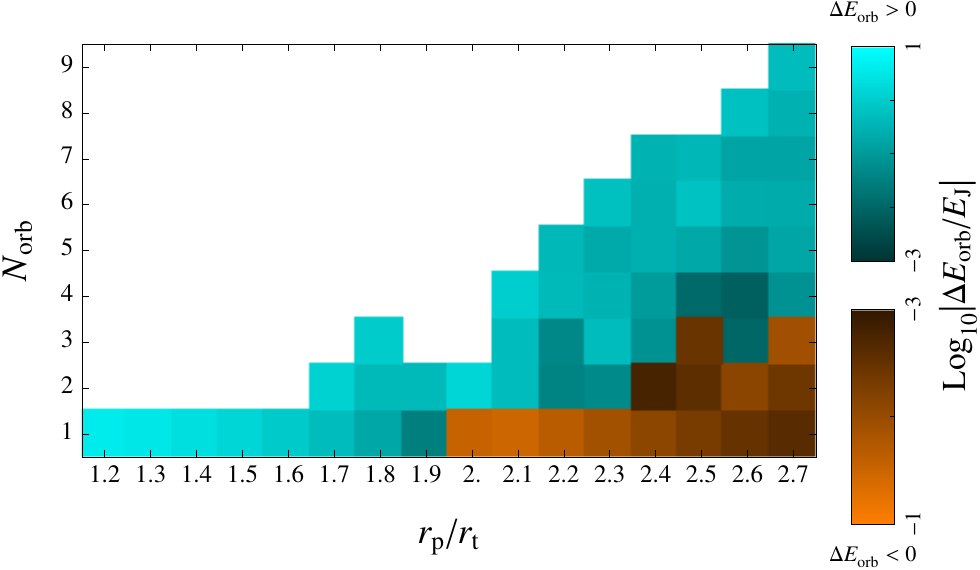}
\caption{Change in orbital energy $E_{\rm orb}$ attributed to each passage as a function of $r_{\rm p}$ and orbit number $N_{\rm orb}$. The orange regions show decreases in $E_{\rm orb}$ (more bound), while the cyan regions show increases in $E_{\rm orb}$ (less bound). The height of each $r_{\rm p}$ column shows the number of orbits a planet survives before being destroyed.}
\label{fig:numorbs}
\end{figure}

This allows us to use the change in orbital energy and angular momentum from our simulations to calculate the orbital evolution for orbits with semi-major axes of several au. The change in orbital energy as a result of each encounter is shown in Figure \ref{fig:numorbs}. For grazing encounters with little mass loss, the change in orbital energy is negative; the planet becomes progressively more bound to the star after each encounter. The magnitude of this change is related to the state of the dynamical tide on the planet at pericenter, where the interaction between the oscillation of the fundamental modes can interact with the tidal field to reduce the amount of energy removed from the orbit, or even change its sign (see Section \ref{sec:chaos}). As the mass loss per orbit exceeds $\sim 10\%$, the trend in orbital energy change becomes positive, and planets become progressively less bound on each subsequent encounter. This is primarily a result of the asymmetrical mass loss, although the interaction with the normal modes is important for encounters where the mass loss is on the order of a few percent.

\begin{figure}[t]
\centering\includegraphics[width=\linewidth,clip=true]{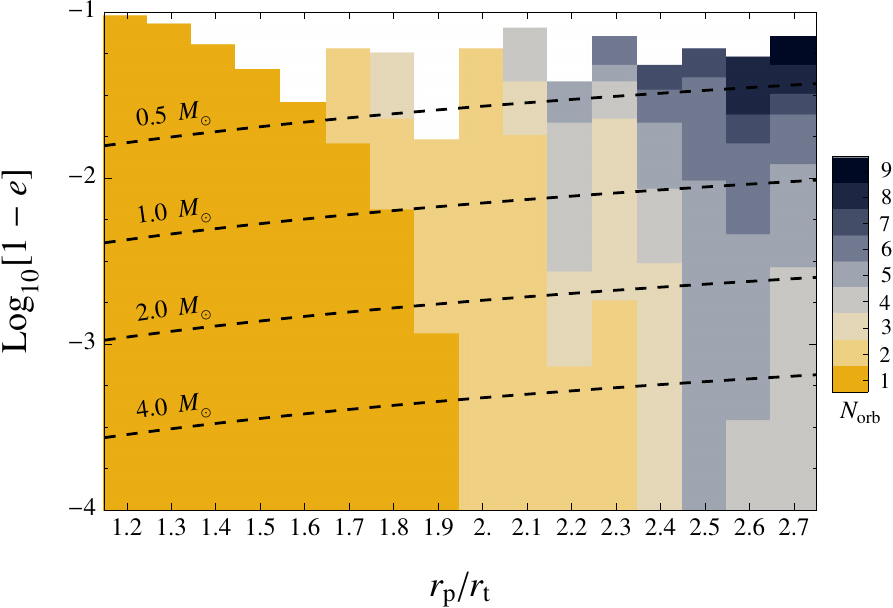}
\caption{Criteria leading to planet ejection given an initial periastron passage distance $r_{\rm p}$ and eccentricity $e$. The colored regions correspond to the number of orbits before a planet is ejected by its interaction with the parent star. The dashed lines show the value of $e$ given an apocenter distance equal to the ice line $a_{\rm ice}$, assuming that $L_{\ast} \propto M_{\ast}^{3.9}$. For all values of $r_{\rm p}$ shown and for $e \lesssim 0.97$, planets are ejected before they are totally disrupted. Note that the region bounding the planets that are ejected on the first orbit has a monotonic dependence on $r_{\rm p}$, while all other regions exhibit more complicated structure. This is because the first passage involves a planet with no internal motions, and thus no relation between phase and the excitation or de-excitation of fundamental modes. All future passages for $N > 1$ involve a planet that is both differentially rotating and oscillating, resulting in a large variance in the amount of energy added or removed from the orbit.}
\label{fig:ejection}
\end{figure}

For orbits with $a \sim a_{\rm ice}$, where $a_{\rm ice} = 2.7 (L_{\ast}/L_{\odot})^{1/2}$ au \citep{Ida:2008p4416} is the ice line, the total orbital energy is small relative to the self-binding energy of the planet, and a positive $\Delta E$ can lead to the surviving planet becoming unbound from the host star. The number of orbits completed by a planet before it becomes unbound is shown in Figure \ref{fig:ejection}. All Jupiter-like planets that come within $r_{\rm p} \leq 2.7 r_{\rm t}$ are expected to be ejected if $e \gtrsim 0.97$, and destroyed otherwise. For the most grazing encounters and smallest eccentricities, we find that the planet can survive for as many as ten orbits before being destroyed by the host star. The general trend is that planets that come closer to their parent stars tend to be ejected or destroyed in fewer orbits, but the consequences of the encounter seem to depend heavily on the orientation of the planet's long axis at pericenter, which as we will show in the next section is quite difficult to predict.

\subsection{The Role of Chaos}\label{sec:chaos}
As the planet continues in its orbit after its first encounter with the parent star, the planet exhibits both rotation and oscillation. Both the magnitude and sign of the orbital energy change is related to the phase of the planet's dynamical tide at periastron. Because the tidal forces strongly excite the $l = 2$ modes, the coherence or decoherence of these modes at periastron can greatly change the effects of that particular encounter \citep{Kochanek:1992p4680, Mardling:1995p4495, Usami:1997p4407}. As the ratio of the orbital period to the break-up rotation period for a Jupiter-like planet scattered in from the ice line is $\sim (M_{\rm J}/M_{\ast})^{1/2}(a_{\rm ice}/R_{\rm J})^{3/2} \sim 10^{5}$, even a very small change in the orbital parameters introduces a dramatic variability in the amount mass lost and the change in orbital energy.

\begin{figure}[tb]
\centering\includegraphics[width=\linewidth,clip=true]{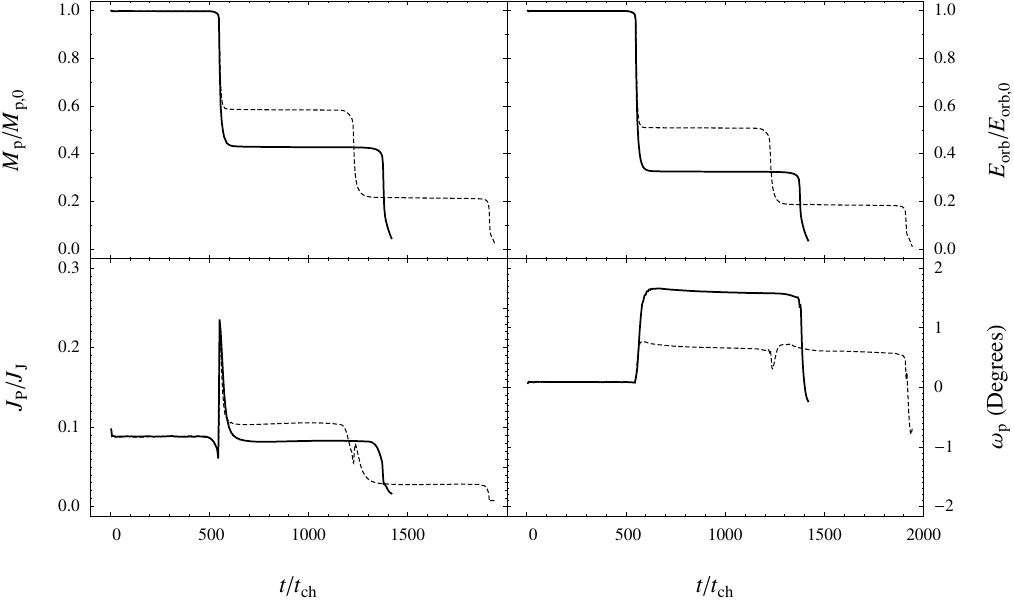}
\caption{The evolution of two multiple encounter simulations with almost identical initial conditions. Both simulations use the same initial conditions as the rest of the multiple encounter simulations presented in this paper, with $r_{\rm p} = 2r_{\rm t}$. The only difference between the two simulations is the initial eccentricity: The solid curves show the outcome for initial eccentricity $e = 0.9$, while the dotted curves show the outcome for $e = 0.90012$, which corresponds to an orbital period that is one free-fall time-scale \(t_{\rm ff}\) longer than the $e = 0.9$ case. The four plots show the planetary mass $M_{\rm P}$, orbital energy $E_{\rm orb}$, spin angular momentum $J_{\rm P}$, and precession of the orbit $\omega$ as functions of $t/t_{\rm ch}$. Note that while the outcome of the first encounter at $t = 0$ is almost completely identical in both simulations, the behavior diverges on the second encounter, which leads to the planet on the $e = 0.9$ orbit being destroyed in three orbits, whereas the planet on the $e = 0.90012$ is destroyed in four. This divergence is a result of the phase difference between the two simulations introduced by the slight difference in initial orbital period.}
\label{fig:chaos}
\end{figure}

For our simulations where $e = 0.9$, the planet completes hundreds revolutions between periastron passages, which means extremely fine sampling of $r_{\rm p}$ would be required to completely describe the problem. To explore the chaotic behavior we ran another multiple disruption simulation with $r_{\rm p} = 2 r_{\rm t}$, with the only difference being that the initial eccentricity is set to $e = 0.90012$, corresponding to an orbital period that is one free-fall timescale \(t_{\rm ff} \equiv (R_{\rm J}^{3}/GM_{\rm J})^{1/2}/2\pi\) longer than the corresponding $e = 0.9$ simulation. As shown in Figure \ref{fig:chaos}, multiple outcomes are possible for even slight changes in the initial conditions, with the planet surviving for a different number of orbits in the two simulations. The chaotic behavior is also evident for our multiple disruption simulation where $r_{\rm p} = 1.8 r_{\rm t}$, which is destroyed on its fourth periastron passage, whereas both the $r_{\rm p} = 1.7 r_{\rm t}$ and $r_{\rm p} = 1.9 r_{\rm t}$ runs are destroyed on their third passages. This chaotic behavior arises because the angle of the major axis of the oscillating planetary core can range from being perpendicular to parallel to the true anomaly at periastron, which dramatically affects the ability of a planet to survive on any particular orbit.

As the principle component of the tidal field are the $l = 2, m = \pm 2$ harmonics, the only way to avoid chaotic behavior is to somehow remove energy from these modes before the planet returns to periastron. For fully-convective, Jupiter-like planets, this can be potentially achieved through three types of mechanisms: Viscosity (either microscopic or turbulent), coupling to other oscillatory modes, or the increase in entropy associated with sound waves steepening into shocks at either the surface of the planet or within its interior.  It should be emphasized that most of the work investigating these energy-sharing mechanisms have concentrated on systems where the oscillations can be treated linearly, and thus the results of these studies can only give us a rough idea to their importance when applied to partially-disruptive encounters.

The two main differences between the linear models and the survivors of a partially-disruptive encounter are the amplitude of the oscillations, which are close to unity, and the presence of a hot, optically thick envelope that accumulates after the disruption and sits on top of the oscillating core. We know that in order for a particular mechanism to result in significant damping of the $l = 2$ modes, the damping timescale has to be at least on the order of the orbital period $P$, if not shorter. As all of the mechanisms for removal of energy from the $l = 2$ modes result in a cascade to microscopic scales, systems with effective damping will experience inflation of the core, inflation of the envelope, or inflation of both regions. This inevitably leads to reduced survivability on subsequent passages.

We now briefly discuss the applicability and viability of each of these proposed mechanisms. For Jupiter-like planets, the microscopic and turbulent fluid viscosities seem to be too small to produce any significant damping on an orbital timescale \citep{Guillot:2004p5120}. Perhaps a more promising mechanism is the coupling of the primary $l = 2$ modes to higher-order ``daughter'' modes, which then couple to ``grand-daughter'' modes, etc., in a cascade resembling the cascade of energy from large scales to small scales in turbulent fluids \citep{Kumar:1996p4773}. In the linear regime, the fundamental mode normally couples to the low frequency g-modes, with the degree of coupling being related to the amplitude of the primary perturbation relative to the size of the object, $\Delta R / R$. But for Jupiter-like planets, which are fully convective and have a negligible luminosity, the polytropic index $\Gamma$ is equal to the adiabatic index $\gamma$, which makes their interiors incapable of supporting g-modes \citep{Cowling:1941p4799}. Coupling can still occur through p-modes, which have higher frequencies than the fundamental mode, but the coupling is only effective for large displacements where the behavior becomes non-linear and for which the rate of energy-sharing is highly uncertain \citep{Kumar:1996p4773}.

Inertial waves may be an effective means of dissipating the $l = 2$ modes given low-frequency tidal forcing \citep{Ivanov:2010p5136}, but the efficiency of this dissipation as the forcing frequency approaches the characteristic frequency is not well understood. And because inertial waves are most effective when the planet spin frequency is comparable to the orbital frequency, they may not be important during the first few passages before syncronicity is established. Additional dissipation may occur via the interaction between inertial waves and Hough waves \citep{Ogilvie:2004p5111}, in which effective coupling is achieved when the wavelength of the inertial modes is comparable to the size of the radiative zone. As scattered planets tend to have large eccentricities, stellar insolation is unlikely to be important for these planets prior to circularization, but as the hot envelope produced by partially disruptive events is radiative, effective dissipation via Hough waves may still be possible.

In our simulations, we do not find that the $l = 2$ modes decay appreciably between periastron passages, despite the fact that the planet oscillates thousands of times in the course of each orbit. We do find, however, that the re-accretion of loosely-bound material moderates the chaotic behavior somewhat. As the density in this region is $\gtrsim 10^3$ times smaller than the core, the tidal radius for the hot envelope component is $\gtrsim 10$ larger than the core's initial tidal radius, which results in the hot envelope being easily removed on subsequent encounters. And unlike the core, the hot envelope has a large sound-crossing time relative to the periastron passage time, and thus no fundamental modes that could affect the interaction on future passages are excited in this region. This leads to the result that as the envelope becomes a larger fraction of the planet's total mass, the behavior becomes notably less chaotic.

\subsection{Debris Accreted by the Star}\label{sec:debacc}
\begin{figure}[tb]
\centering\includegraphics[width=\linewidth,clip=true]{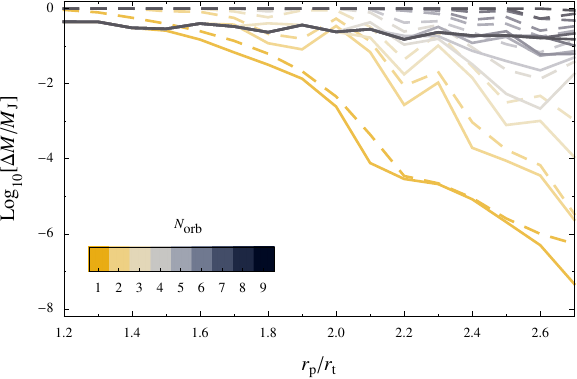}
\caption{Mass loss history of multiple passage encounters for different initial values of $r_{\rm p}$. Each curve is color-coded to correspond to a particular orbit number. The solid lines show the aggregate mass accreted by the star, while the dashed lines show the total mass lost from the planet.}
\label{fig:massloss}
\end{figure}
The amount of mass removed on the first few orbits can vary over many orders of magnitude for a relatively small range of $r_{\rm p}$. For our closest simulated encounters, nearly half the planet ends up being bound to the host star after the first passage, while our most grazing encounters show almost no mass loss at all (Figure \ref{fig:massloss}). However, as the initial passages excite fundamental modes of oscillation, and the returning debris leads to additional heating in the planet's outer layers, the amount of mass removed geometrically increases with each subsequent passage. This leads to the result that approximately 20-50\% of the planet's initial mass is accreted by the host star by the time the planet has been completely disrupted, with slightly less material being accreted when the initial encounter is grazing.

The reason closer disruptions lead to more mass accreted by the central star has to do with the asymmetry of the tides; the force resulting from the inner tide (closest to the star) at pericenter is
\begin{equation}
    \frac{\left(2 r_{\rm p} - R_{\rm P}\right)\left(R_{\rm P}+r_{\rm p}\right)^2}{\left(2r_{\rm p}+R_{\rm P}\right)\left(R_{\rm P} - r_{\rm p}\right)^2} \simeq 1 + \frac{3 R_{\rm P}}{r_{\rm p}} {\rm\;as\;} r_{\rm p} \rightarrow \infty
    \label{eq:forceratio}
\end{equation}
times larger than the outer tide (furthest from the star). And while the effective size of the planet increases due to the excitation of oscillations and the loss of mass, subsequent passages yield diminishing returns as material has already been removed on previous encounters.

As the mass lost by the planet from a series of partially disruptive encounters is highly stochastic, the exact amount of mass accreted by the star for any given multiple-orbit encounter is difficult to predict. To parameterize the average mass accreted by the star, we consider two limiting cases. If the planet is scattered into an orbit with $a \ll a_{\rm ice}$, the planet will be completely destroyed before it is ejected. On the other hand, if a planet is scattered from the ice line, $E_{\rm orb}$ is much smaller than the planet's binding energy and the planet will be ejected on the orbit for which $\Delta E_{\rm orb} > E_{\rm orb}$. A fit of $\Delta M_\ast$ for these two limiting cases yields
\begin{equation}
    \Delta M_\ast (\beta) = \left\{
    \begin{array}{rl}
        1.26\,\exp \left[-0.79 \beta^{-1}\right]M_{\rm J} & : a_0 \ll a_{\rm ice}\\
        \smallskip\\
        9.62\,\exp \left[-2.59 \beta^{-1}\right]M_{\rm J} & : a_0 \sim a_{\rm ice},
    \end{array}
    \right.
    \label{eq:macc}
\end{equation}
for $0.37 \leq \beta \leq 0.83$, where the impact parameter $\beta \equiv r_{\rm t} / r_{\rm p}$. For the case where the planet is completely destroyed, the amount of mass accreted is nearly constant, with a slight decrease with decreasing $\beta$. The decrease is substantially steeper for the incomplete disruptions, as planets on grazing orbits are less likely to transfer much mass to their parent stars prior to being ejected.

\section{Discussion}\label{sec:disc}
\subsection{The Jupiter Exclusion Zone}
\begin{figure*}[t]
\centering\includegraphics[width=\linewidth,clip=true]{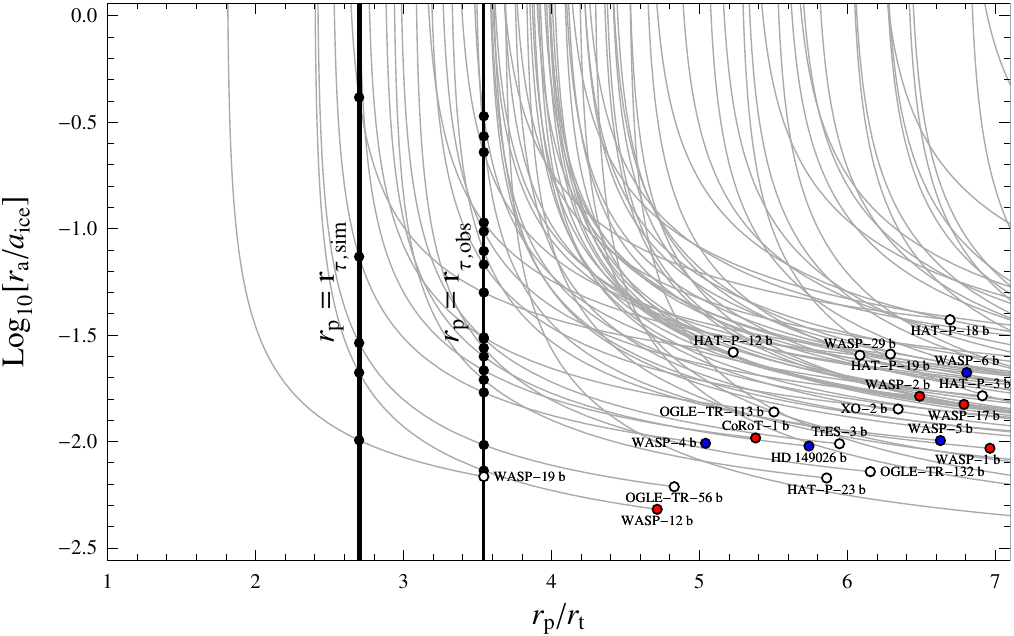}
\caption{Possible apastrons $r_{\rm a}$ for the known hot Jupiters with $M_{\rm P} > 0.1 M_{\rm J}$. Each arc shows the apocenter of an orbit with pericenter $r_{\rm p}$, assuming that the angular momentum of the initial orbit is equal to the orbital angular momentum observed today. The open circles show each planet's currently observed $r_{\rm p}$ and $r_{\rm a}$, scaled to the tidal radius $r_{\rm t}$ and the ice line $a_{\rm ice}$ respectively, which are determined by the planet/host star properties. Blue-filled circles are planets that are thought to be aligned with their host stars, red-filled circles are thought to be misaligned \citep[either via direct measurement or because they have observed to have statistically inconsistent rotation rates, see][]{Schlaufman:2010p4689}, and white-filled circles show systems with unknown orientations. The thick vertical line (labeled $r_{\tau, {\rm sim}}$) shows the minimum possible value of $r_\tau$ corresponding to the Jupiter exclusion zone, which is determined through our numerical simulations for Jupiter-like planets to be $2.7r_{\rm t}$, while the thinner vertical line (labeled $r_{\tau, {\rm obs}}$) shows the maximum possible value for $r_{\tau}$, which is defined by the planet that is currently observed to be closest to its classically defined tidal radius, WASP-19 b. Filled black circles show the intersection between the $r_{\tau}$ lines and each of the arcs of constant angular momentum shows the maximum apastron distance $r_{\rm a, \max}$ a planet could have been scattered from without being destroyed on its subsequent encounters with the host star. Note that the typical $r_{\rm a, \max}$ values are significantly smaller than $a_{\rm ice}$, which indicates that those planets must have migrated prior to being scattered if planet-planet scattering brought them to their current positions. All data taken from the Extrasolar Planets Encyclopaedia (\url{http://exoplanet.eu}) and Ren\'{e} Heller's Holt-Rossiter-McLaughlin Encyclopaedia (\url{http://www.aip.de/People/RHeller/}) on December 8, 2010.}
\label{fig:maxa}
\end{figure*}
As discussed in Section \ref{sec:simres}, there exists an exclusion zone with radius $r_{\tau} = 2.7 r_{\rm t}$ within which all Jupiter-like planets are either ejected or destroyed. For our simulations we used a polytropic model of a gas giant, with a mass and radius equal to present-day Jupiter. As some gas giant planets may not have had much time to cool before being disrupted, it is likely that disrupted Jupiter-like planets are larger than our fiducial cold Jupiter model \citep{Bodenheimer:2001p5115}. This is confirmed by observations, many of the known hot Jupiters are observed to be inflated, with larger radii and lower densities as a result of the injection of entropy via stellar insolation \citep{Fortney:2007p4838, Miller:2009p4910, Li:2010p4958}, and potentially also other dissipative mechanisms \citep{Ogilvie:2004p5111, Laine:2009p5799, Arras:2010p5142}. But this means that the tidal radii for these planets must be {\it larger} than what we use in our calculations, translating to planets that are more easily disrupted than cold gas giants. And while the core mass of Jupiter-like planets is uncertain, all plausible models include cores that compose such a small fraction of the planet's total mass that they are unimportant in determining the structure of the gas envelope, and thus the dynamics of the disruption. The value of $r_\tau$ calculated in this paper is thus a lower limit for Jupiter-like planets. For gas giants with masses more similar to Neptune, the affect of an increased core mass may allow these planets to survive closer to their parent stars due to the increase in average density. We stress that our model should only be applied to planets of $M \gtrsim 0.1 M_{\rm J}$.

Because the spatial resolution required to resolve grazing passages becomes computationally prohibitive beyond $\sim 3 r_{\rm t}$, the true value of $r_\tau$ may lie beyond what we have been able to calculate for even cold Jupiter-like planets. Thus, the vertical cut-off line denoted by $r_{\tau, {\rm sim}}$ in Figure \ref{fig:maxa} almost certainly lies closer to the right of the diagram, which would add more planets to the list that could not have been directly scattered to their present positions from the ice line. However, the true cutoff value certainly lies no further than the current pericenter distance of WASP-19 b (denoted as $r_{\tau, {\rm obs}}$), currently the exoplanet with the smallest known value of $r_{\rm p}/r_{\rm t}$, which appears to be quite inflated \citep{Hebb:2010p5125} and probably has a tidal radius that is larger than a cold gas giant of the same mass. Hence, we can only constrain $r_\tau$ to lie within the range $r_{\tau, {\rm sim}} \leq r_\tau \leq r_{\tau, {\rm obs}}$, or $2.7 \leq r_\tau/r_{\rm t} \leq 3.54$.

\subsection{Implications for the Formation of Hot Jupiters}\label{sec:implications}
\begin{deluxetable*}{ccccc}
    \tablecolumns{5}
    \tablehead{\colhead{Planet} & \colhead{$r_{\rm p}/r_{\rm t}$} & \colhead{$r_{a,\max}/a_{\rm ice}$} & \colhead{$Q_{\ast,\max}$\tablenotemark{a}} & $\tau_{{\rm life}, \max}$\tablenotemark{a} \\
    & & & & \colhead{(Gyr)}}
    \startdata
        HAT-P-12 b   & 5.2 & 0.41  & $ 1 \times 10^4$ --- $1 \times 10^5$ & 0.4  --- 4\\
        OGLE-TR-56 b & 4.8 & 0.029 & $ 2 \times 10^7$ --- $4 \times 10^7$ & 0.9  --- 2\\
        WASP-4 b     & 5.1 & 0.093 & $ 5 \times 10^7$ --- $2 \times 10^8$ & 1    --- 6\\
        WASP-12 b    & 4.7 & 0.021 & $ 5 \times 10^7$ --- $2 \times 10^8$ & 0.5  --- 1\\
        WASP-19 b    & 3.5 & 0.010 & $ 2 \times 10^7$ --- $2 \times 10^8$ & 0.07 --- 1
    \enddata
    \tablenotetext{a}{Assuming $r_{\rm p,0} = 2 r_{\tau}$, $a_0 = a_{\rm ice}$. The minimum and maximum values are calculated using the lower and upper limits for $\tau_{\rm age}$ as compiled by \cite{Schlaufman:2010p4689}.}
    \label{tab:planets}
\end{deluxetable*}

Because specific orbital angular momentum is very nearly conserved during even strong tidal encounters (Figure \ref{fig:trajectories}), the currently observed semi-major axes of hot Jupiters is at most 2 $r_{\rm p, 0}$, where $r_{\rm p, 0}$ is the planet's initial pericenter distance after being scattered into a disruptive orbit. As the exclusion radius $r_{\tau}$ is larger than the $a/2$ for many known exoplanets, there exists a maximum initial apastron distance from which a planet could have been scattered without having been destroyed or ejected
\begin{equation}
r_{\rm a, max} = \frac{a_{\rm obs}\left(1-e_{\rm obs}^{2}\right)r_{\tau}}{a_{\rm obs}\left(e_{\rm obs}^{2} - 1\right) + 2r_{\tau}},
\end{equation}
where $a_{\rm obs}$ and $e_{\rm obs}$ are the currently observed semi-major axis and eccentricity. This maximum apastron value is illustrated for the currently known hot Jupiters in Figure \ref{fig:maxa}. Because a planet that is scattered into a disruptive orbit from the ice line possesses $e \sim 1$, the maximum allowed apastron is a highly sensitive function of $r_{\tau}$ for $r_{\rm a} \gtrsim 0.1 a_{\rm ice}$. Thus, if the current distance of an exoplanet from its host star is less than $2 r_{\tau}$, it is highly likely that its initial orbit prior to scattering must have been significantly closer to the star than the ice line. This means that no Jupiter-like planets would be observed for initial pericenter distances smaller than this value if they were scattered from the ice line to their present-day orbits.

Five of the currently known hot Jupiters (HAT-P-12 b, OGLE-TR-56 b, WASP-4 b, WASP-12 b, and WASP-19 b) have observed semi-major axes $a_{\rm obs}>2r_{\tau, {\rm sim}}$, corresponding to $r_{\rm a,\max} < a_{\rm ice}$ (Table \ref{tab:planets}). For four of the five planets, the maximum initial apastron distance is less than 0.1 $a_{\rm ice}$. As all Jupiter-mass planets are thought to form beyond $a_{\rm ice}$, these planets could not have been {\it directly} scattered to their current locations. This implies that a migration process that resulted in a dramatic decrease in $a$ must have occurred either before or after these planets were scattered. If the true value of $r_{\tau}$ for cold Jupiter-like planets lies closer to the maximum possible value set by WASP-19 b ($r_{\tau, {\rm obs}}$), as many as 18 of the currently known hot Jupiters with $M > 0.1 M_{\rm J}$ could not have survived the scattering event that deposited them at their current locations.

As both planet-planet scattering and the Kozai mechanism do not radically alter a planet's semi-major axis, disk migration would be required to explain how close-in planets would have initial apastrons that are significantly smaller than $a_{\rm ice}$ \citep{Lin:1996p5522, Ida:2004p4433, Ida:2008p4416}. As many of the currently known exoplanets have $r_{a,\max} < a_{\rm ice}$, this would imply that the migration timescale in these systems must have been substantially shorter than the disk lifetime if the planets migrated prior to the scattering event. For the misaligned systems, a dynamical process (either planet-planet scattering or the Kozai mechanism) may need to occur after the migration or while the gas disk dissipates \citep{Matsumura:2010p4877} to produce the observed misalignment.

Alternatively, the planets may have migrated after the scattering event. For this to be the case, the scattering event had to bring the planet close enough to its parent star such that the tide raised on the star can alter the planet's orbit in a time shorter than the system age, but not too close that the planet is destroyed or ejected by the star. Suppose that a planet is scattered from beyond $a_{\rm ice}$ such that $r_{\rm p,0} > r_\tau$. After circularization the pericenter distance doubles to $2 r_\tau$, and the planet's semi-major axis evolves via the interaction between the planet and the tide it raises on the surface of the star \citep{Eggleton:1998p5222,Hut:1980p5217}. At this distance, the orbital period of the planet is almost always shorter than the spin-period of the star, except perhaps for stars with ages $\lesssim 650$ Myr \citep{Irwin:2009p5489}. This results in a spin-up of the star, which tries to ``catch up'' to the Keplerian frequency imposed by the planet's orbit, and an inward migration of the planet.

As the planet is likely to be tidally locked even prior to circularization, the timescale for evolution of the planet's semi-major axis is entirely determined by the star's properties and the orbital frequency $\omega$ \citep{DobbsDixon:2004p5198},
\begin{equation}
    \frac{a}{\dot{a}} = \frac{1}{9} Q_{\ast} \left( \frac{M_\ast}{M_{\rm P}} \right) \left( \frac{a}{R_\ast} \right)^5 \left(\omega - \Omega_\ast\right)^{-1},
    \label{eq:atimescale}
\end{equation}
where $R_\ast$ is the star's radius, $Q_\ast$ is the star's tidal quality factor, and $\Omega_\ast$ is its rotation frequency. The fastest inward migration occurs when a star is not rotating, e.g. $\Omega_\ast \rightarrow 0$. As planet-planet scattering seems to be rare after $10^8$ yr \citep{Matsumura:2008p4494}, we can set $a/\dot{a}$ equal to the system age $\tau_{\rm age}$ to determine $Q_{\ast,\max}$, the maximum tidal quality factor for which a planet can migrate from $2 r_\tau$ to $a_{\rm obs}$
\begin{align}
    Q_{\ast, \max} &= 7 \times 10^5 \left( \frac{M_{\rm P}}{M_{\rm J}} \right)^{8/3} \left( \frac{M_\ast}{M_\odot} \right)^{-8/3} \left( \frac{R_{\rm P}}{R_{\rm J}} \right)^{-5}\nonumber\\
    &\times \left( \frac{r_\tau}{3 r_{\rm t}} \right)^{-5} \left( \frac{P_0}{3\;{\rm days}} \right)^{-1} \left( \frac{\tau_{\rm age}}{ \rm Gyr} \right),
    \label{eq:qmax}
\end{align}
where $P_0$ is the initial orbital period. When setting $r_\tau = r_{\tau, {\rm sim}}$, all of the known hot Jupiters with $a_{\rm obs} < 2 r_{\tau}$ yields values for $Q_{\ast,\max}$ that are consistent with those expected for stars (Table \ref{tab:planets}). 

If the true radius of disruption lies closer to $r_{\tau, {\rm obs}}$, the measured values of $Q_{\ast,\max}$ are more restrictive, meaning that higher dissipation rates would be required to enable planet-planet scattering to viably produce hot Jupiters with $a < 2 r_\tau$. This would also imply that most observed hot Jupiters would only exist a short while longer before being destroyed by their parent star. This remaining lifetime can be calculated by solving Equation (\ref{eq:atimescale}) using present-day values for $a$, $\omega$, and $\Omega_\ast$ (denoted $\tau_{\rm life, \max}$ in Table \ref{tab:planets}). As $Q_{\ast,\max}$ is extremely sensitive to the exact value of $r_{\tau}$, only a slight increase in $r_\tau$ is required to eliminate direct scattering as a possible genesis mechanism for a significant fraction of the hot Jupiters. If it can be shown that $r_\tau$ is substantially larger than what we have calculated in this work (i.e. $r_\tau \simeq r_{\tau, {\rm obs}}$), $Q_{\ast,\max}$ can put definitive constraints on the mechanism responsible for producing hot Jupiters.

\subsection{Stellar Spin-Up from Planetary Disruption}\label{sec:spinup}
\begin{figure*}[tb]
\centering\includegraphics[width=\linewidth,clip=true]{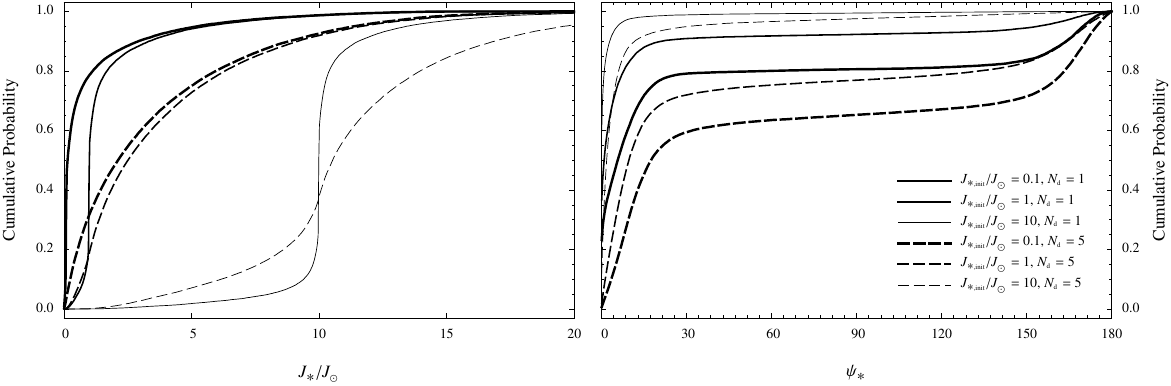}
\caption{Changes to the stellar spin as a result of accreting tidally disrupted planets. The curves show the cumulative probability that a star will possess a given angular momentum $J_{\ast}$ and spin inclination $\psi_{\ast}$ after 1 (solid) or 5 (dashed) planetary disruptions $N_{\rm d}$ for different initial stellar angular momentum $J_{\rm\ast, init}$. Planets are assumed to have a logarithmic distribution in mass ($0.1 M_{\rm J} \le M \le 10 M_{\rm J}$) and semi-major axes $(r_\tau \le a \le 10 a_{\rm ice})$, where $r_\tau$ is the minimum distance for which a planet won't be tidally disrupted and $a_{\rm ice}$ is the ice line. The eccentricity $e$ and inclination $i$ relative to the invariable plane are assumed to follow Rayleigh distributions, with $\sigma_{e} = 0.3$ and $\sigma_{i} = 10^\circ$.}
\label{fig:spinup}
\end{figure*}

As discussed in Section \ref{sec:debacc}, the star acquires a substantial injection of angular momentum from the disrupted planet. Our simulations give us an empirical determination of \(\Delta M_\ast\left(\beta, M_{\rm P} = M_{\rm J}, M_\ast = 10^3 M_{\rm J}\right)\), the amount of mass gained by a star with $M_\ast = 10^3 M_{\rm J}$ from the disruption of a Jupiter-mass planet given the impact parameter $\beta$ (Figure \ref{fig:massloss}). The angular momentum acquired by the star is simply equal to the specific angular momentum at the star's radius multiplied by the amount of mass accreted, $\Delta J_\ast = \Delta M_\ast \sqrt{GM_\ast R_\ast}$. Note that this is not the same as the total angular momentum content of the disk formed from the debris as claimed by \cite{Jackson:2009p4413}, as the accretion stream does not intersect the star's radius unless the planet's original orbit has $r_{\rm p} < R_\ast$ \citep{Kochanek:1994p671}.

If all disrupted Jupiter-like planets are approximately the same size and $e \sim 1$, all disruptions can be treated using our fiducial model and the inclusion of the simple scaling associated with the increase in the orbital angular momentum. This is because the only difference that arises from changing the mass of the planet and the star is the degree of asymmetry of the tides, which is third-order in the force expansion and is $(R_{\rm J} / \beta r_{\rm t})^2 \sim 10^{-2}$ times smaller than the tidal force itself. Thus, in the event that the planet is completely destroyed by the star, we expect that the amount of angular momentum acquired by the star should simply scale with $M_\ast$ and $M_{\rm P}$
\begin{align}
    \Delta J_\ast(\beta, M_{\rm P}, M_\ast) &= \left(\frac{M_\ast}{M_{\odot}}\right)^{1/2} \left(\frac{M_{\rm P}}{M_{\rm J}}\right)^{3/2}\nonumber\\
    &\times \Delta J_\ast(\beta, M_{\rm J}, 10^3 M_{\rm J}).
    \label{eq:deltaj}
\end{align}

However, our results show that a substantial fraction of planets are ejected from the system before they can be fully destroyed by the star. This reduces the amount of mass $\Delta M_\ast$ accreted by the star, and thus $\Delta J_\ast$. Therefore $\Delta J_\ast$ for any given encounter should depend on which particular orbit a planet is ejected $N_{\rm ej}$. This parameter depends on the orbital energy $E_{\rm orb}$, and the change in orbital energy associated with each encounter, which as we explain in Section \ref{sec:multpass} should also should scale simply with the masses of the planet and the star. Thus our expression for $\Delta J_\ast$ becomes
\begin{align}
    \Delta J_\ast(\beta, M_{\rm P}, M_\ast) &= \left(\frac{M_\ast}{M_{\odot}}\right)^{1/2} \left(\frac{M_{\rm P}}{M_{\rm J}}\right)^{3/2}\nonumber\\
    &\times \Delta J_\ast(\beta, M_{\rm J}, 10^3 M_{\rm J}, N_{\rm ej})\label{eq:deltaj2}\\
    \Sigma_{N = 1}^{N_{\rm ej}} \Delta E_{\rm orb} &> -E_{\rm orb},\label{eq:nej}
\end{align}
where $N_{\rm ej}$ is equal to the lowest value of $N$ for which Equation (\ref{eq:nej}) is satisfied. While this gives us the amount of angular momentum acquired by the star for a set of encounter parameters, we must know the distribution of $\beta$ in order to estimate the probability for which a star will possess a total angular momentum $J_\ast$ and obliquity $\psi_\ast$ after a number of planetary disruptions $N_{\rm d}$. 

The integrated rate of scattering (i.e., all encounters with $\beta$ greater than some value) has been evaluated numerically by a number of authors. \cite{Juric:2008p4408} show that up to 20\% of planets present at the end of phase I of planetary formation can collide with the host star, e.g. $r_{\rm p} < R_{\ast}$. \cite{Ford:2008p4328} show that up to 16\% of planets in a 3-body system can be thrown into a ``star-grazing'' orbit, which they define as $r_{\rm p} < 10^{-2} a_{\rm init}$. \cite{Nagasawa:2008p4405} show that planet-planet scattering events tend to induce transitions between Kozai states (with the duration of each state being $\sim 10^{6} - 10^{7}$ yr) until a planet is ejected from the system or until the eccentricity of the innermost planet is damped by tides. \citeauthor{Nagasawa:2008p4405} note that most of the close-in planets seem to be driven to their closest approaches via the Kozai mechanism, which gently drives the planets into the region where they can be circularized, as opposed to being directly scattered into such orbits. None of the models presented above include the precession associated with general relativity that would normally quench the Kozai mechanism.

Because the numerical scattering experiments do not include a hydrodynamical treatment of tides, the fates of planets that are either scattered or driven by the Kozai to $r_{\rm t} \leq r_{\rm p} \leq r_{\tau}$ are not accurately represented. What the scattering experiments {\it do} reveal is the total rate of planet-planet interactions as a function of the number of gravitating bodies in the system, and the distribution of orbits that arises from numerous planet-planet interactions. And despite the simplistic treatment of tides or the neglect of tides altogether, different prescriptions of tidal dissipation do not seem to strongly affect the distribution of the remaining planets in the system \citep{Nagasawa:2008p4405}, which means that the orbit distributions these models predict should still be appropriate to use as inputs for our post-disruption stellar spin estimates.

For systems that are not dynamically stable after the gas disk dissipates, or for systems which are driven to instability by an external perturber \citep{Zhou:2007p5956, Malmberg:2010p4876}, the models indicate that the eccentricity distribution of planets quickly evolves to a Rayleigh distribution \citep{Juric:2008p4408}
\begin{equation}
dN = \frac{e}{\sigma_{e}^{2}}\exp\left(\frac{-e^{2}}{2 \sigma_{e}^{2}}\right) de
\label{eq:eccdist}
\end{equation}
As the relaxation to this distribution is mainly driven by strong two-body planet-planet interactions, the new eccentricity $e$ of a planet after a scattering event should also be drawn from the above distribution. Because we are interested in objects that may be unbound from the system after the encounter, objects with $e$ initially larger than 1 should be included when calculating the number of events at each $e$. Using our definition of $\beta$,
\begin{equation}
e = 1 - \frac{2}{\beta r_{\rm t} / r_{\rm a} + 1},
\end{equation}
and by making a change of variable from $e$ to $\beta$, Equation (\ref{eq:eccdist}) becomes
\begin{equation}
dN = \frac{2 r_{\rm a} r_{\rm t} (r_{\rm a} \beta - r_{\rm t})}{(r_{\rm t}+r_{\rm a} \beta )^3 \sigma_{e}^2}\exp\left[-\frac{(r_{\rm t}-r_{\rm a} \beta )^2}{2 (r_{\rm t}+r_{\rm a} \beta )^2 \sigma_{e}^2}\right] d\beta.
\label{eq:fulldn}
\end{equation}
When $r_{\rm a} \gg r_{\rm p}$, as is the case in planet disruption, the above expression simplifies to
\begin{equation}
dN \simeq \frac{2 r_{\rm t}}{r_{\rm a}\sigma_{e}^{2}}\exp\left[-\frac{1}{2\sigma_{e}^{2}}\right]\frac{d\beta}{\beta^{2}},
\end{equation}
which when integrated yields a disruption probability that scales inversely with $\beta$ \citep{Rees:1988p9}. Because we are including hyperbolic encounters, $\beta$ can assume both positive and negative values, with $r_{\rm t}/r_{\rm a} < \left|\beta\right| < \infty$. To first order, the value of the integrand is equal for both positive and negative values of $\beta$, and the total number of events where $\beta > \beta^{\prime}$ is given by
\begin{equation}
N = 2\int_{\beta^{\prime}}^{\infty} \frac{dN}{d\beta}d\beta.
\end{equation}
This implies that equal numbers of planets will be scattered into prograde and retrograde orbits. Under these assumptions and given our empirically determined lower limit for ejection $r_{\rm p} = 2.7 r_{\rm t}$, it is immediately clear that the rate of collisions with the central star is lower than the combined rate of ejections and disruptions by at least a factor $r_\tau/R_\ast - 1 = 1.78$, assuming solar and Jovian values.

Now that we have a model for the expected initial distribution of giant planets in a dynamically relaxed system, we can use Equation (\ref{eq:fulldn}) to evaluate the expected values of $J_\ast$ and $\psi_\ast$ after $N_{\rm d}$ planetary disruptions for a star of $M_\ast = M_\odot$ (Figure \ref{fig:spinup}). Here we consider the scattered objects to be cold, Jupiter-like planets with $M_{\rm P} > 0.1 M_{\rm J}$, resulting in a nearly-constant planetary radius $R_{\rm P}$. This assumption is only valid if the system is older than $\sim 10^8$ yr and if the planet's initial orbit is far enough from its parent star to have negligible insolation, but note that including these effects would only act to increase the amount of mass removed from the planet per orbit. We also assume that the planets are distributed uniformly in log $a$ from $r_\tau$ to $10 a_{\rm ice}$, and in log $M_{\rm P}$ from 0.1 to 10 $M_{\rm J}$, with Rayleigh distributions in $e$ and $i$ with $\sigma_e = 0.3$ and $\sigma_i = 10^\circ$. For simplicity, we assume that any angular momentum acquired by the star through a disruption is shared equally with all parts of the star.

For a single planet disruption in a system where the star possesses initial angular momentum equal to the Sun, $\psi_\ast$ exceeds 30$^\circ$ in 15\% of the stellar hosts, and 90$^\circ$ (i.e. the star rotates retrograde to the invariable plane) in 8\% of systems. The spin rate of the star also tends to increase, with 10\% of stars possessing $3J_\odot$ after the disruption. If disruptions are very common ($N = 5$), the probability of $>10/>90^\circ$ increases to 47/22\%, and 41\% of stars have $J_\ast > 3J_\odot$. The probability for enhanced values of $J_\ast$ and $\psi_\ast$ are all slightly smaller if we restrict disruptions to come from $a > a_{\rm ice}$, as the amount of mass the star acquires from disruptions is slightly less on average, Equation (\ref{eq:macc}). However, the effect is minor, with changes in the cumulative probabilities being on the order of a few percent.

If a star is unable to share the angular momentum deposited in its outer layers in a time less than its age, the star may be observed to have larger values of $\Omega_\ast$ or $\psi_\ast$ than what would be expected given complete mixing. The timescale $\tau_\nu$ for sharing angular momentum across the tachocline in the Sun is known to be only $\sim$3 Myr \citep{Gough:1998p5644}, with the timescale decreasing for increasing rotation rates in the convective region. The timescale for sharing angular momentum across the tachocline appears to increase significantly as the size of the convective zone shrinks for rapidly rotating stars \citep{Barnes:2003p5660, Barnes:2010p5664}, but this may be moderated by a fingering instability made possible by the larger molecular weight of the disruption debris \citep{Garaud:2010p5958,Rosenblum:2010p5959}. A disruption in a system where the host star never has a thick convective zone could effectively erase the star's original spin rate and obliquity, with $\psi_\ast$ being pulled from the inclination distribution of planetary orbits. 

\subsection{Observational Signatures}\label{sec:obsig}
If we assume that the pressure profile of the hot atmosphere that accumulates on the planet has a pressure profile $P = P_{\rm ram}$ at all radii \citep{Frank:2002p4864}, the initial Kelvin-Helmholtz cooling timescale given a total atmosphere mass $M_{\rm atm}$ is approximately
\begin{align}
    \tau_{\rm KH} \simeq 1.2 \left(\frac{M_{\rm atm}}{0.1 M_{\rm J}}\right)&\left(\frac{M_{\rm P}}{M_{\rm J}}\right)\left(\frac{R_{\rm P}}{R_{\rm J}}\right)^{-1}\nonumber\\
    \times &\left(\frac{R_{\rm atm}}{3 R_{\rm J}}\right)^{-2}\left(\frac{T_{\rm vir}}{10^5 {\rm \,K}}\right)^{-4} {\,\rm days},
\end{align}
where $T_{\rm vir}$ is the virial temperature, $M_{\rm atm}$ is the mass of the atmosphere and $R_{\rm atm}$ is the radius of the atmosphere. Because $T_{\rm vir}$ is initially very large, the atmosphere is at first completely ionized. At this temperature, $\tau_{\rm KH}$ is a few days, leading to a rapid thermal evolution of the planet's outer layers shortly after most of the sundered mass returns to the planet. During this phase, the planet can briefly outshine its parent star with $L_{\rm bol} \sim 10^{36}$ erg s$^{-1}$, with most of the radiation being emitted in the UV. As the atmosphere cools it shrinks back down onto the planet's surface until $R_{\rm atm} \sim R_{\rm P}$. We can then estimate the temperature at which the planet radiates for the majority of its orbit by setting $P$ equal to $\tau_{\rm KH}$, which yields a temperature of a few $10^4$ K. This indicates that the planet's outer layers will still be somewhat inflated before the planet returns to pericenter, and thus will be easily removed on subsequent passages.

For planets that are ejected from their host stars, the thermal evolution of their outer layers continues until the temperature reaches a few thousand Kelvin, at which point hydrogen begins to recombine, which acts as a thermostat to maintain the temperature at a relatively constant value. The recombination timescale is
\begin{align}
    \tau_{\rm rec} \simeq 150 \left(\frac{X_{\rm H}}{0.7}\right)\left(\frac{M_{\rm atm}}{0.1 M_{\rm J}}\right)\left(\frac{R_{\rm P}}{R_{\rm J}}\right) {\,\rm years},
    \label{}
\end{align}
where $X_{\rm H}$ is the Hydrogen fraction. As a result of this relatively long time-scale, these ejected planets could remain quite bright ($L_{\rm bol} \sim 0.1 L_\odot$) for an extended period of time, even without additional tidal forcing. If we assume that one in ten planet-hosting stars ejects a Jupiter-like planet via a partial disruption, and adopting an average star formation rate of $1 M_\odot$ per year, at least one ejected planet in the recombination phase should be visible in the galaxy at any one time.

\section{Conclusions}\label{sec:conclusions}
\subsection{Limitations and Future Directions}
The principle assumptions that we have made in this paper is that Jupiter-like planets are represented accurately by a polytropic model of its structure. One advantage of this model is that disruption simulations are trivially scalable to planets of a different size by a simple correction to $\beta$, assuming that the planet's mass interior to a given radius $M_{\rm P}(<r)$ scales self-similarly and that the fluid $\gamma$ remains unchanged. An $n = 1$ polytrope reproduces the mass profile of coreless 1 $M_{\rm J}$ planet relatively well, with the difference in $M_{\rm P}(<r)$ never exceeding 10\% throughout the planet's interior (N. Miller, private communication).

The inclusion of a core of a few tens of $M_\oplus$ affects $M_{\rm P}(<r)$ out to a few times the core radius, for which $M_{\rm P}(<r) \sim 0.4 M(R_{\rm P})$. Beyond this radius, the structure of the planet is nearly identical to the coreless/polytropic models. This means that our simulations should be an accurate representation for disruptions where $\lesssim 70\%$ of the planet's mass is removed for Jupiter-like planets. For Neptune-like planets, where the core mass can be larger than the gas mass, the difference in $M_{\rm P}(<r)$ is substantial all the way to the planet's outermost layers, and thus our simulation results should not be directly applied. As the average densities of Neptune-like planets is larger than Jupiter-like planets, $r_\tau$ for Neptunes should assume a smaller value.

Additionally, we assume that $\gamma = 2$ throughout the simulation volume, even for regions of very low density where the fluid is completely ionized and should behave as an ideal gas ($\gamma = 5/3$) or even a radiation pressure dominated fluid ($\gamma = 4/3$) in the lowest-density regions. This transition to different values of $\gamma$ should affect the structure of the hot envelope that forms from the re-accreted debris that surrounds a partially disrupted planet, which is dynamically unimportant but may affect the planet's observable signature. This is not to say that a more realistic equation of state would not affect the mass loss itself. As the process of ripping material from the planet involves rapid fluid decompression, a decrease in $\gamma$ may result in slightly altered disruption dynamics.

Ideally, one would like to extend the models we have presented here to include a more physical equation of state that can treat all components of the pre- and post-disrupted planet realistically. As the resolution required to determine $r_\tau$ for multiple-orbit encounters beyond what we have presented here is prohibitive, it seems that the exploring the affects of using a more-complete equation of state with a realistic initial planet model is the next natural step for future studies. In the case of planets with a substantial core, these modifications are necessary to determine $r_\tau$ with any confidence.

\subsection{The Fates of Scattered Jupiters}
The fate of a Jupiter-like planet after a strong scattering event is a function of the strength of the tidal forces it experiences at periastron. In this paper, we have determined the disruption radius $r_\tau$ for Jupiter-like planets which sets the boundary between long-term survival and rapid tidal disruption. Below, we summarize the various post-scattering outcomes in order of decreasing distance, using $r_\tau$ and the tidal radius $r_{\rm t}$ as points of reference.

Stalled $\left(r_{\rm p} \gtrsim 6 r_{\rm t}\right)$: The planet is deposited into an orbit where the rate of tidal dissipation is too small to result in a change in semi-major axis over the lifetime of the system. This planet may be in a Kozai state driven by a third body in the system, or could experience another strong scattering event, which may lead to an increase of eccentricity and subsequent circularization.

Circularization/Migration $\left(r_\tau < r_{\rm p} < 6 r_{\rm t}, e \lesssim 0.9\right)$: In this region, the planet is close enough to its parent star that tidal dissipation is effective, and the planet can circularize in $10^9$ yr or less for moderate values of $e$. For near-radial orbits, circularization may still be longer than the stellar age, but again the Kozai mechanism or scattering could lead to a more rapid orbital evolution. All currently observed hot Jupiters are either stalled, in the process of circularizing, or already have circular orbits. If the planet is close enough to its parent star and $Q_\ast \lesssim 10^7$, the planet will raise a tide on the star and migrate inwards due to the transfer of angular momentum.

Ejection $\left(r_{\rm p} < r_\tau, e \gtrsim 0.97\right)$: A planet that passes within the exclusion zone will be ejected from the system if its initial orbit is radial enough such that its orbital energy $E_{\rm orb}$ is significantly smaller than the self-binding energy $E_{\rm p}$ of the planet. Slightly less than half of the planet's initial mass remains bound to the central star, carrying with it a large reservoir of angular momentum that can significantly alter the host star's spin rate and axis of rotation. The ejected planet will remain as bright or brighter than its host star for a few years, eventually plateauing via hydrogen recombination as an object with a fraction of solar luminosity for a century. All Jupiter-like planets that scatter in from beyond $a_{\rm ice}$ such that $r_{\rm p} < r_\tau$ will be ejected from the system.

Complete disruption $\left(r_{\rm p} < r_\tau, e \lesssim 0.97\right)$: For planets that are deep within their parent star's potential well, the planet cannot soak the change in energy required to significantly alter the orbit, which eventually leads to its complete disruption. Approximately half of the planet's mass accretes onto the stellar host, carrying the same specific angular momentum as in the ejection case, leading to even more pronounced effects on the stellar spin. Only planets that have migrated close to their stars prior to being scattered are destroyed before they are ejected.

Collision with the central star $\left(r_{\rm p} < R_\ast\right)$: The planet strikes the surface of the star directly. Anywhere from half to all of the planet's mass is absorbed by the star, with the angular momentum being carried by this material potentially being smaller than that carried by the debris from a disruption, depending on how direct the impact is. These events are approximately twice as uncommon as ejections/disruptions.

\acknowledgments We have benefited from many useful discussions with E. Ford, K. Schlaufman, E. Quataert, F. Rasio, G. Laughlin, N. Miller, D. Fabrycky, B. Hansen, M. Rees, and A. Socrates. The software used in this work was in part developed by the DOE-supported ASCI/Alliance Center for Astrophysical Thermonuclear Flashes at the University of Chicago. Computations were performed on the Pleiades and Laozi UCSC computer clusters. We acknowledge support from the David and Lucille Packard Foundation (JG and ER-R), NSF grants PHY-0503584 (ER-R) and AST-0908807 (DL), NASA grants NNX07A-L13G (DL), NNX07AI88G (DL), NNX08AL41G (DL), and NNX08AM84G (DL); and the NASA Earth and Space Science Fellowship (JG).
\appendix
\section{Modified PPM Gravity Algorithm}\label{sec:modgrav}
\begin{figure*}[t]
\centering\includegraphics[width=\linewidth,clip=true]{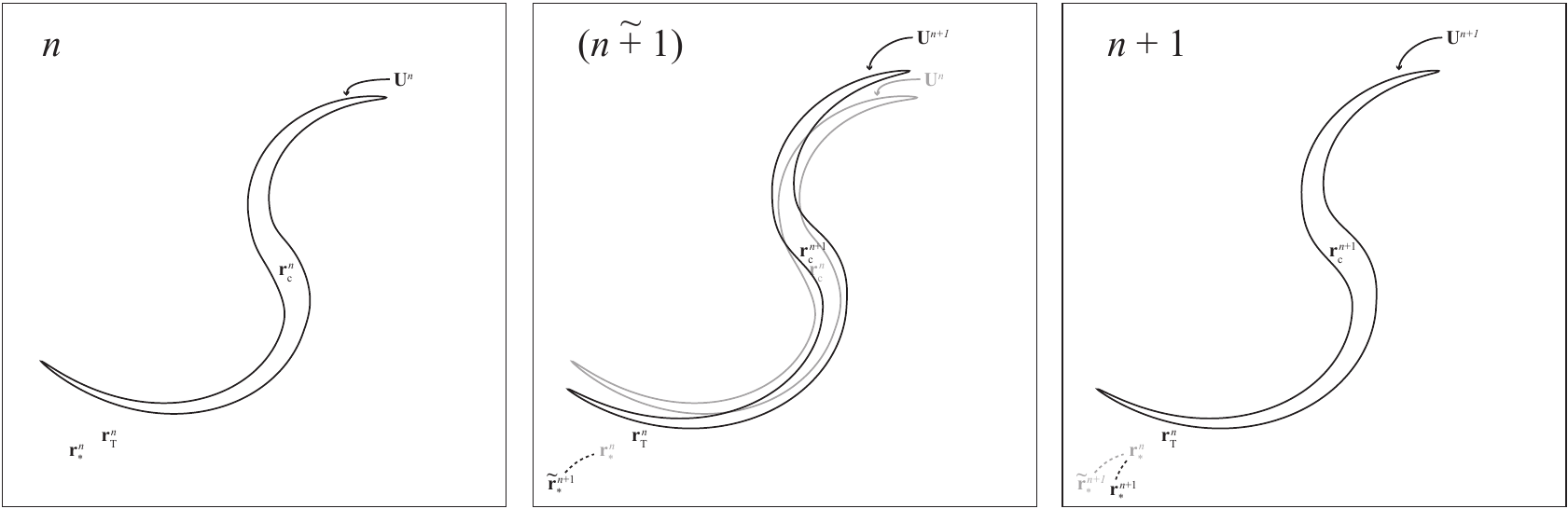}
\caption{Cartoon showing the sequence of states used to compute the evolution over a single a time-step using our modified gravity solver. The initial state $m$ is showing in the left panel, the middle panel shows the intermediate state $\tilde{(m+1)}$, and the last panel shows the final state $m + 1$. $\mathbf{r}_{\ast}$ is the position of the point mass representing the star, $\mathbf{r}_{\rm c}$ is the planetary core's true center of mass, $\mathbf{r}_{\rm T} \equiv (M_{\rm P} \mathbf{r}_{t} + M_{\ast} \mathbf{r}_{\ast}) / (M_{\rm J} + M_{\ast})$ is the center of mass of the complete system. The variable vector field $\mathbf{U}(\mathbf{r}) \equiv [\rho, \rho \mathbf{v}, e, X]$ represents the value of all the conserved quantities in a given state. Differences between the states are exaggerated for illustrative purposes.}
\label{fig:cartoon}
\end{figure*}

Because the binding energy of a planet on a disruptive orbit is comparable to the planet self-binding energy, the conservative properties of a code used to investigate planetary disruption are important. As the simulation of a partially-disruptive encounter involves the simultaneous resolution of both a compact core and two debris tails which are hundreds of time larger than the core, we found that the standard methods used to calculate the gravitational potential in a tidal disruption are too computationally expensive given the required accuracies.

Our approach was to improve upon the gravity solver found within the FLASH hydrodynamics code \citep{Fryxell:2000p440} such that it is better suited to investigating the problem of tidal disruption, a pictorial representation of the algorithm described below is shown in Figure \ref{fig:cartoon}. Two centers of mass are calculated at each time-step, which are defined as
\begin{align}
\mathbf{r}_{\rm t} &= \frac{\sum_{i} m_{i} r_{i}}{\sum_{i} m_{i}}\\
\mathbf{r}_{\rm c} &= \frac{\sum_{i,\,\rho\left(i\right) > f\rho_{\max}} m_{i} r_{i}}{\sum_{i,\,\rho\left(i\right) > f\rho_{\max}} m_{i}},
\end{align}
where $\mathbf{r}_{\rm t}$ is the ``true'' center of mass and $\mathbf{r}_{\rm c}$ is the ``core'' center of mass, which only includes matter above a density cut-off $\rho > f\rho_{\max}$, where $f$ is set to $0.1$. The planet's virtual particle is fixed to spatially coincide with the $\mathbf{r}_{\rm c}$ vector at all times. The planet's self-gravity is calculated using a multipole expansion of the planet's mass about $\mathbf{r}_{\rm c}$ instead of $\mathbf{r}_{\rm t}$, which allows us to better approximate the planet's potential using less terms in the multipole expansion.

Our algorithm is particularly well-suited for investigating tidal disruptions. If the expansion were performed about $\mathbf{r}_{\rm t}$ after large tidal tails have formed within the simulation, $\mathbf{r}_{\rm t}$ would be mostly determined by the position of material that is far away from the core, and the region with the best resolution of the potential may lie in empty space. We can estimate the number of terms required to represent the core's potential if the expansion is carried out about $\mathbf{r}_{\rm t}$ instead of $\mathbf{r}_{\rm c}$. Assume the core has a radius $r_{\rm c}$, and that the core lies a distance $d \equiv |\mathbf{r}-\mathbf{r}_{\rm c}|$ from the true center of mass. The angular scale of a lobe corresponding to a spherical harmonic of degree $l$ is simply $\pi/l$, which means that even a first-order approximation of the core's potential requires an expansion with $l_{\max} \ge \pi d / r_{\rm c}$. Assuming $\sim 10\%$ of a planet's mass is lost during an encounter and this material lies an average distance $\sim 10^{3} R_{\rm J}$ from the core at apocenter, $d$ is on order $100 R_{\rm J}$, meaning that the multipole expansion must be carried out to $l \gtrsim 300$. This is highly impractical, and thus it is much more efficient to carry out the multipole expansion about the planet's core, whose position is associated with the densest material in the simulation and is where the potential gradients are largest. 

The potential $\phi$ calculated from the multipole expansion about $\mathbf{r}_{\rm c}$ is used both to apply forces to the fluid in the simulation domain and to the virtual star and planet particles. A consequence of not expanding about the true center of mass is that there exists a non-zero force that is applied to the core. These forces are associated with the odd-$l$ multipole terms that usually cancel when the expansion is carried out about the true center of mass. Our multipole expansion does not discard these odd-$l$ terms, which allows us to confidently represent the fluid's potential using an expansion of relatively low order.

In the FLASH code's split PPM formalism, the equations used for coupling hydrodynamics and the gravitational field for a cell $i$ along each of the three cartesian directions are \citep{Bryan:1995p4683} 
\begin{align}
    \left(\rho v\right)_{i}^{m+1} &= \left(\rho v\right)_{i}^{m} + \frac{\Delta t^{m}}{2} g_{i}^{m+1}\left(\rho_{i}^{m} + \rho_{i}^{m+1}\right)\label{eq:hyd1}\\
    \left(\rho E\right)_{i}^{m+1} &= \left(\rho E\right)_{i}^{m} + \frac{\Delta t^{m}}{4} g_{i}^{m+1}\left(\rho_{i}^{m} + \rho_{i}^{m+1}\right)\left(v_{i}^{m} + v_{i}^{m+1}\right)\label{eq:hyd2}.
\end{align}
The acceleration $g_{i}^{m+1}$ is calculated by extrapolating $\phi_{i}^{m-1}$ and $\phi_{i}^{m}$ to obtain an estimate for $\phi_{i}^{m+1}$
\begin{equation}
\tilde{\phi}_{i}^{m+1} = \phi_{i}^{m}\left(1 + \frac{\Delta t^{m}}{\Delta t^{m-1}}\right) - \phi_{i}^{m-1}\frac{\Delta t^{m}}{\Delta t^{m-1}}.
\label{eq:phi1}
\end{equation}
This is in turn used to calculate $g_{i}^{m+1}$
\begin{equation}
    g_{i}^{m+1} = \frac{1}{2 \Delta x_{i}} \Bigl[\tilde{\phi}_{i+1}^{m+1} - \phi_{i-1}^{m+1}+\frac{1}{12}\left(\tilde{\phi}_{i+1}^{m+1} - 2\phi_{i}^{m+1} + \phi_{i - 1}^{m+1} \frac{\delta \rho_{i}}{\rho_{i}}\right)\Bigr],
\label{eq:gi}
\end{equation}
where $\delta\rho_{i}$ is defined as
\begin{equation}
\delta\rho_{i} = \min\left(\left|\rho_{i + 1}^{m} - \rho_{i - 1}^{m}\right|, 2\left|\rho_{i}^{m}-\rho_{i-1}^{m}\right|, 2\left|\rho_{i}^{m}-\rho_{i+1}^{m}\right|\right)\times\rm{sign\left(\rho_{i+1}^{m} - \rho_{i-1}^{m}\right)}
\end{equation}
to enforce monotonicity in $\rho$. 

Because the potential of the star in our simulations is approximated by an analytical expression (for our simulations, a monopole potential), we can implement the following modification such that the component of the acceleration attributed to the star can be calculated to much higher accuracy. We first make the assumption that the matter distribution remains fixed over the course of the time-step, and then integrate the virtual particle positions forward in time from $m$ to $m + 1$. This allows us to calculate
\begin{equation}
    \tilde{\phi}_{\ast,i}^{m+1} = \frac{G M_{\ast} \rho^{m}_{ {\rm P},i}}{\left|\mathbf{r}_{i} - \mathbf{r}^{m+1}_{\ast}\right|},
\end{equation}
an estimate of the star's contribution to the potential based on the star's approximate final position. This estimate should be much closer to the true value than simple extrapolation as the potential at a given location has a non-trivial time-dependence. Splitting $\phi_i$ into two components $\phi_{\ast, i}$ (star) and $\phi_{ {\rm P}, i}$ (planet), Equation (\ref{eq:phi1}) becomes
\begin{equation}
    \tilde{\phi}_{i}^{m+1} = \phi_{{\rm P},i}^{m}\left(1 + \frac{\Delta t^{m}}{\Delta t^{m-1}}\right) - \phi_{ {\rm P},i}^{m-1}\frac{\Delta t^{m}}{\Delta t^{m-1}} + \tilde{\phi}_{\ast,i}^{m+1}.
\end{equation}
Additionally, a correction must be made to Equation (\ref{eq:gi})
\begin{equation}
    g_{i,{\rm cor}}^{m+1} = g_{i}^{m+1} - g_{\rm c}^{m+1},
    \label{eq:gcor}
\end{equation}
where $g_{\rm c}^{m+1}$ is the acceleration experienced by the core due to the presence of odd-$l$ terms in the multipole expansion. Using $g_{i,{\rm cor}}^{m+1}$ instead of $g_i^{m+1}$, a hydrodynamical step is performed according to Equations (\ref{eq:hyd1}) and (\ref{eq:hyd2}), which yields $\rho_i^{m+1}$ and thus the true contribution at $m + 1$ of the planet to the potential, $\phi_{ {\rm P}, i}^{m+1}$. The positions of the virtual particles are then re-integrated from $m$ to $m + 1$, but this time using a linear interpolation of the time-evolving potential over the time-step.

\begin{figure*}[t]
\centering\includegraphics[width=\linewidth,clip=true]{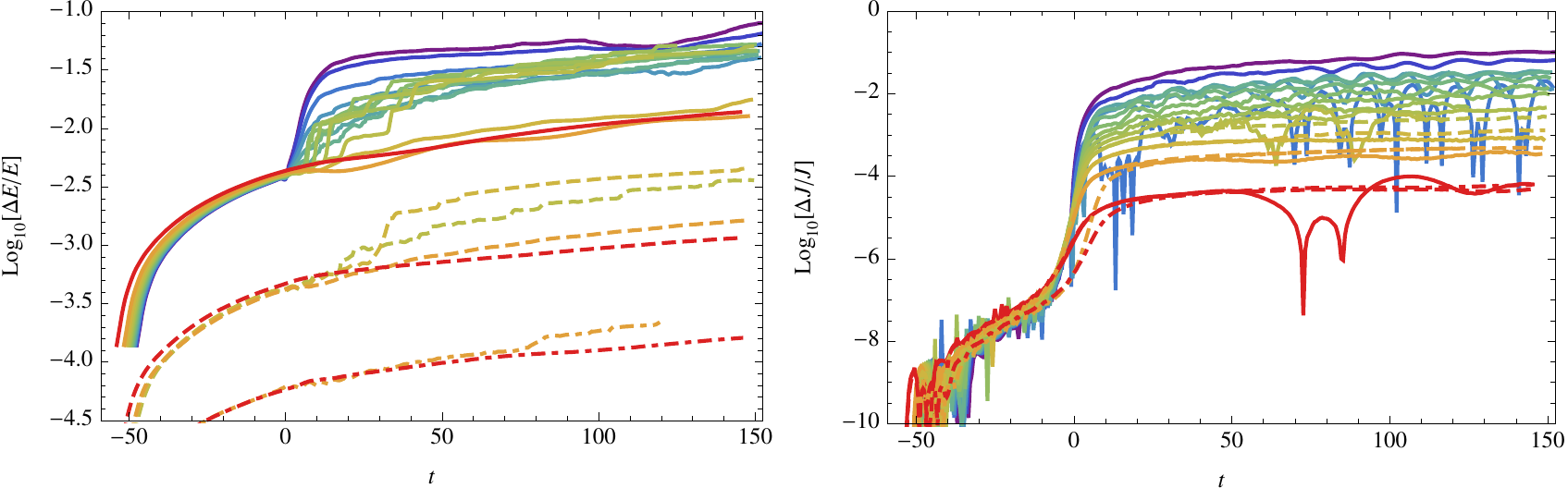}
\caption{Accumulation of numerical error in the total energy $E$ and angular momentum $J$ for the simulations ran to compare with the results of \citetalias{Faber:2005p4315}. The solid lines show simulations for which $s = 0.02 R_{\rm J}$, the dashed lines show simulations where $s = 0.01 R_{\rm J}$, and the dash-dotted lines show simulations where $s = 5 \times 10^{-3} R_{\rm J}$. Note that the rate of error accumulation is greatest shortly after periastron when the planet's self-potential is rapidly varying as a function of time. As the planet returns to a state of quasi-equilibrium, the rate of error accumulation asymptotes to the original rate.}
\label{fig:conserve}
\end{figure*}

As mentioned previously, a complication introduced by this method is that the net force applied to the point about which the multipole expansion is carried out is non-zero. This is because we do not cancel out the acceleration applied to the extended object's true center of mass $\mathbf{r}_{\rm t}$, but rather the center of mass defined by the densest material, $\mathbf{r}_{\rm c}$. While our correction to the gravitational acceleration (Equation (\ref{eq:gcor})) mitigates the problem somewhat, the mass density within the simulation domain varies as a function of time, and thus the total mass that satisfies the density criteria to be included when calculating $\mathbf{r}_{\rm c}$ also changes as a function of time. This leads to small, but measurable, changes in the reference point over the course of a simulation which are not entirely corrected by simply subtracting off the force applied at $\mathbf{r}_{\rm c}$. To ensure that angular momentum and energy are conserved, a correction is made to virtual particle positions over the course each time-step when re-integrating from $m$ to $m + 1$,
\begin{equation}
    \mathbf{x}_{\ast,cor}\left(t\right) = \mathbf{x}_{\ast}\left(t\right) - \frac{t - t^{m}}{t^{m+1}-t^{m}} \left(\mathbf{r}_{\rm c}^{m+1} - \mathbf{r}_{\rm c}^{m}\right).
\end{equation}
In practice, $\mathbf{r}_{\rm c}^m$ and $\mathbf{r}_{\rm c}^{m+1}$ are very nearly equal, with a substantial difference only arising when the extended object is close to being destroyed. Typical corrections are significantly smaller than the size of individual grid cells in the most highly-resolved regions. 

The accumulation of error for our single passage encounters is shown in Figure \ref{fig:conserve}. In the deepest encounters where the self-potential changes rapidly, we find that the rate of relative error accumulation in total energy and angular momentum is no larger than $10^{-4}$ per dynamical time for our simulations with the lowest maximum-resolution, with the planet's initial diameter being resolved by 50 grid cells at $t = 0$. Most of this error arises from the ejected tidal tails, which are resolved at lower resolution out of necessity because of the large volume they occupy. For encounters in which all the matter within the simulation is always resolved at highest resolution (i.e. those that do not have significant mass loss), the relative error accumulation is only $10^{-6}$ per dynamical time.

We also tested the error accumulation for two sets of simulations with higher maximum resolutions, with the planet's initial diameter being resolved by 100 and 200 grid cells. These simulations show an accumulation of fractional error of $\sim 10^{-7}$ and $10^{-8}$ per dynamical time, respectively.

\bibliographystyle{apj}
\bibliography{apj-jour,2009d}

\begin{thebibliography}{77}
\expandafter\ifx\csname natexlab\endcsname\relax\def\natexlab#1{#1}\fi

\bibitem[{Antonini {et~al.}(2010)Antonini, Lombardi, \&
  Merritt}]{Antonini:2010p4875}
Antonini, F., Lombardi, J., \& Merritt, D. 2010, eprint arXiv:1008.5369

\bibitem[{Arras \& Socrates(2010)}]{Arras:2010p5142}
Arras, P. \& Socrates, A. 2010, \apj, 714, 1

\bibitem[{Barnes(2003)}]{Barnes:2003p5660}
Barnes, S.~A. 2003, \apj, 586, 464

\bibitem[{Barnes \& Kim(2010)}]{Barnes:2010p5664}
Barnes, S.~A. \& Kim, Y.-C. 2010, \apj, 721, 675

\bibitem[{Bicknell \& Gingold(1983)}]{Bicknell:1983p604}
Bicknell, G.~V. \& Gingold, R.~A. 1983, \apj, 273, 749

\bibitem[{Bodenheimer {et~al.}(2001)Bodenheimer, Lin, \&
  Mardling}]{Bodenheimer:2001p5115}
Bodenheimer, P., Lin, D. N.~C., \& Mardling, R.~A. 2001, \apj, 548, 466

\bibitem[{Bryan {et~al.}(1995)Bryan, Norman, Stone, Cen, \&
  Ostriker}]{Bryan:1995p4683}
Bryan, G.~L., Norman, M.~L., Stone, J.~M., Cen, R., \& Ostriker, J.~P. 1995,
  Computer Physics Communications, 89, 149

\bibitem[{Carter \& Luminet(1983)}]{Carter:1983p12}
Carter, B. \& Luminet, J. 1983, \aap, 121, 97

\bibitem[{Carter \& Luminet(1985)}]{Carter:1985p238}
Carter, B. \& Luminet, J.~P. 1985, \mnras, 212, 23

\bibitem[{Cowling(1941)}]{Cowling:1941p4799}
Cowling, T.~G. 1941, \mnras, 101, 367

\bibitem[{Diener {et~al.}(1997)Diener, Frolov, Khokhlov, Novikov, \&
  Pethick}]{Diener:1997p457}
Diener, P., Frolov, V.~P., Khokhlov, A.~M., Novikov, I.~D., \& Pethick, C.~J.
  1997, \apj, 479, 164

\bibitem[{Dobbs-Dixon {et~al.}(2004)Dobbs-Dixon, Lin, \&
  Mardling}]{DobbsDixon:2004p5198}
Dobbs-Dixon, I., Lin, D. N.~C., \& Mardling, R.~A. 2004, \apj, 610, 464

\bibitem[{Eggleton {et~al.}(1998)Eggleton, Kiseleva, \&
  Hut}]{Eggleton:1998p5222}
Eggleton, P.~P., Kiseleva, L.~G., \& Hut, P. 1998, \apj, 499, 853

\bibitem[{Evans \& Kochanek(1989)}]{Evans:1989p147}
Evans, C.~R. \& Kochanek, C.~S. 1989, \apjl, 346, L13

\bibitem[{Faber {et~al.}(2005)Faber, Rasio, \& Willems}]{Faber:2005p4315}
Faber, J.~A., Rasio, F.~A., \& Willems, B. 2005, Icarus, 175, 248

\bibitem[{Fabrycky \& Tremaine(2007)}]{Fabrycky:2007p5801}
Fabrycky, D. \& Tremaine, S. 2007, \apj, 669, 1298

\bibitem[{Ford \& Rasio(2008)}]{Ford:2008p4328}
Ford, E.~B. \& Rasio, F.~A. 2008, \apj, 686, 621

\bibitem[{Fortney {et~al.}(2007)Fortney, Marley, \& Barnes}]{Fortney:2007p4838}
Fortney, J.~J., Marley, M.~S., \& Barnes, J.~W. 2007, \apj, 659, 1661

\bibitem[{Foucart \& Lai(2010)}]{Foucart:2010p4925}
Foucart, F. \& Lai, D. 2010, arXiv

\bibitem[{Frank {et~al.}(2002)Frank, King, \& Raine}]{Frank:2002p4864}
Frank, J., King, A., \& Raine, D.~J. 2002, Accretion Power in Astrophysics

\bibitem[{Frolov {et~al.}(1994)Frolov, Khokhlov, Novikov, \&
  Pethick}]{Frolov:1994p2}
Frolov, V.~P., Khokhlov, A.~M., Novikov, I.~D., \& Pethick, C.~J. 1994, \apj,
  432, 680

\bibitem[{Fryxell {et~al.}(2000)Fryxell, Olson, Ricker, Timmes, Zingale, Lamb,
  MacNeice, Rosner, Truran, \& Tufo}]{Fryxell:2000p440}
Fryxell, B., Olson, K., Ricker, P., Timmes, F.~X., Zingale, M., Lamb, D.~Q.,
  MacNeice, P., Rosner, R., Truran, J.~W., \& Tufo, H. 2000, \apj, 131, 273

\bibitem[{Garaud(2010)}]{Garaud:2010p5958}
Garaud, P. 2010, arXiv

\bibitem[{Gough \& McIntyre(1998)}]{Gough:1998p5644}
Gough, D.~O. \& McIntyre, M.~E. 1998, \nat, 394, 755

\bibitem[{Guillochon {et~al.}(2009)Guillochon, Ramirez-Ruiz, Rosswog, \&
  Kasen}]{Guillochon:2009p3441}
Guillochon, J., Ramirez-Ruiz, E., Rosswog, S., \& Kasen, D. 2009, \apj, 705,
  844

\bibitem[{Guillot {et~al.}(2004)Guillot, Stevenson, Hubbard, \&
  Saumon}]{Guillot:2004p5120}
Guillot, T., Stevenson, D.~J., Hubbard, W.~B., \& Saumon, D. 2004, In: Jupiter.
  The planet, 35

\bibitem[{Hebb {et~al.}(2010)Hebb, Collier-Cameron, Triaud, Lister, Smalley,
  Maxted, Hellier, Anderson, Pollacco, Gillon, Queloz, West, Bentley, Enoch,
  Haswell, Horne, Mayor, Pepe, Segransan, Skillen, Udry, \&
  Wheatley}]{Hebb:2010p5125}
Hebb, L., Collier-Cameron, A., Triaud, A., Lister, T., Smalley, B., Maxted, P.,
  Hellier, C., Anderson, D., Pollacco, D., Gillon, M., Queloz, D., West, R.,
  Bentley, S., Enoch, B., Haswell, C., Horne, K., Mayor, M., Pepe, F.,
  Segransan, D., Skillen, I., Udry, S., \& Wheatley, P. 2010, \apj, 708, 224

\bibitem[{Hubbard(1984)}]{Hubbard:1984p5843}
Hubbard, W.~B. 1984, New York

\bibitem[{Hut(1980)}]{Hut:1980p5217}
Hut, P. 1980, \aap, 92, 167

\bibitem[{Ida \& Lin(2004)}]{Ida:2004p4433}
Ida, S. \& Lin, D. N.~C. 2004, \apj, 604, 388

\bibitem[{Ida \& Lin(2008)}]{Ida:2008p4416}
---. 2008, \apj, 685, 584

\bibitem[{Irwin \& Bouvier(2009)}]{Irwin:2009p5489}
Irwin, J. \& Bouvier, J. 2009, IAU Symposium, 258, 363

\bibitem[{Ivanov \& Novikov(2001)}]{Ivanov:2001p4490}
Ivanov, P.~B. \& Novikov, I.~D. 2001, \apj, 549, 467

\bibitem[{Ivanov \& Papaloizou(2010)}]{Ivanov:2010p5136}
Ivanov, P.~B. \& Papaloizou, J. C.~B. 2010, \mnras, 407, 1609

\bibitem[{Jackson {et~al.}(2009)Jackson, Barnes, \&
  Greenberg}]{Jackson:2009p4413}
Jackson, B., Barnes, R., \& Greenberg, R. 2009, \apj, 698, 1357

\bibitem[{Juri{\'c} \& Tremaine(2008)}]{Juric:2008p4408}
Juri{\'c}, M. \& Tremaine, S. 2008, \apj, 686, 603

\bibitem[{Khokhlov {et~al.}(1993{\natexlab{a}})Khokhlov, Novikov, \&
  Pethick}]{Khokhlov:1993p5432}
Khokhlov, A., Novikov, I.~D., \& Pethick, C.~J. 1993{\natexlab{a}},
  Astrophysical Journal v.418, 418, 181

\bibitem[{Khokhlov {et~al.}(1993{\natexlab{b}})Khokhlov, Novikov, \&
  Pethick}]{Khokhlov:1993p5430}
---. 1993{\natexlab{b}}, Astrophysical Journal v.418, 418, 163

\bibitem[{Kobayashi {et~al.}(2004)Kobayashi, Laguna, Phinney, \&
  M{\'e}sz{\'a}ros}]{Kobayashi:2004p152}
Kobayashi, S., Laguna, P., Phinney, E.~S., \& M{\'e}sz{\'a}ros, P. 2004, \apj,
  615, 855

\bibitem[{Kochanek(1992)}]{Kochanek:1992p4680}
Kochanek, C.~S. 1992, \apj, 385, 604

\bibitem[{Kochanek(1994)}]{Kochanek:1994p671}
---. 1994, \apj, 422, 508

\bibitem[{Kozai(1962)}]{Kozai:1962p5046}
Kozai, Y. 1962, \aj, 67, 591

\bibitem[{Kumar \& Goodman(1996)}]{Kumar:1996p4773}
Kumar, P. \& Goodman, J. 1996, \apj, 466, 946

\bibitem[{Lai {et~al.}(1994)Lai, Rasio, \& Shapiro}]{Lai:1994p3305}
Lai, D., Rasio, F.~A., \& Shapiro, S.~L. 1994, \apj, 437, 742

\bibitem[{Laine {et~al.}(2009)Laine, Lin, \& Dong}]{Laine:2009p5799}
Laine, R., Lin, D., \& Dong, S. 2009, \apj, 685, 521

\bibitem[{Lee {et~al.}(2010)Lee, Ramirez-Ruiz, \& van~de Ven}]{Lee:2010p4412}
Lee, W.~H., Ramirez-Ruiz, E., \& van~de Ven, G. 2010, \apj, 720, 953

\bibitem[{Li {et~al.}(2010)Li, Miller, Lin, \& Fortney}]{Li:2010p4958}
Li, S.-L., Miller, N., Lin, D. N.~C., \& Fortney, J.~J. 2010, \nat, 463, 1054

\bibitem[{Lin {et~al.}(1996)Lin, Bodenheimer, \& Richardson}]{Lin:1996p5522}
Lin, D. N.~C., Bodenheimer, P., \& Richardson, D.~C. 1996, \nat, 380, 606

\bibitem[{Lodato {et~al.}(2009)Lodato, King, \& Pringle}]{Lodato:2009p3051}
Lodato, G., King, A.~R., \& Pringle, J.~E. 2009, \mnras, 392, 332

\bibitem[{Lor{\'e}n-Aguilar {et~al.}(2009)Lor{\'e}n-Aguilar, Isern, \&
  Garc{\'\i}a-Berro}]{LorenAguilar:2009p4672}
Lor{\'e}n-Aguilar, P., Isern, J., \& Garc{\'\i}a-Berro, E. 2009, \aap, 500,
  1193

\bibitem[{Malmberg {et~al.}(2010)Malmberg, Davies, \&
  Heggie}]{Malmberg:2010p4876}
Malmberg, D., Davies, M.~B., \& Heggie, D.~C. 2010, arXiv, astro-ph.EP

\bibitem[{Mardling(1995)}]{Mardling:1995p4495}
Mardling, R.~A. 1995, \apj, 450, 722

\bibitem[{Matsumura {et~al.}(2008)Matsumura, Takeda, \&
  Rasio}]{Matsumura:2008p4494}
Matsumura, S., Takeda, G., \& Rasio, F.~A. 2008, \apj, 686, L29

\bibitem[{Matsumura {et~al.}(2010)Matsumura, Thommes, Chatterjee, \&
  Rasio}]{Matsumura:2010p4877}
Matsumura, S., Thommes, E.~W., Chatterjee, S., \& Rasio, F.~A. 2010, \apj, 714,
  194

\bibitem[{Miller {et~al.}(2009)Miller, Fortney, \& Jackson}]{Miller:2009p4910}
Miller, N., Fortney, J.~J., \& Jackson, B. 2009, \apj, 702, 1413

\bibitem[{Nagasawa {et~al.}(2008)Nagasawa, Ida, \& Bessho}]{Nagasawa:2008p4405}
Nagasawa, M., Ida, S., \& Bessho, T. 2008, \apj, 678, 498

\bibitem[{Nolthenius \& Katz(1982)}]{Nolthenius:1982p5484}
Nolthenius, R.~A. \& Katz, J.~I. 1982, Astrophysical Journal, 263, 377, a{\&}AA
  ID. AAA032.066.185

\bibitem[{Ogilvie \& Lin(2004)}]{Ogilvie:2004p5111}
Ogilvie, G.~I. \& Lin, D. N.~C. 2004, \apj, 610, 477

\bibitem[{Press {et~al.}(1986)Press, Flannery, \& Teukolsky}]{Press:1986p1001}
Press, W.~H., Flannery, B.~P., \& Teukolsky, S.~A. 1986, Cambridge: University
  Press

\bibitem[{Press \& Teukolsky(1977)}]{Press:1977p3083}
Press, W.~H. \& Teukolsky, S.~A. 1977, \apj, 213, 183

\bibitem[{Ramirez-Ruiz \& Rosswog(2009)}]{RamirezRuiz:2009p3071}
Ramirez-Ruiz, E. \& Rosswog, S. 2009, \apjl, 697, L77

\bibitem[{Rathore(2005)}]{Rathore:2005p4797}
Rathore, Y. 2005, Proquest Dissertations And Theses 2005. Section 0037, 23

\bibitem[{Rees(1988)}]{Rees:1988p9}
Rees, M.~J. 1988, \nat, 333, 523

\bibitem[{Robertson {et~al.}(2009)Robertson, Kravtsov, Gnedin, Abel, \&
  Rudd}]{Robertson:2009p3764}
Robertson, B.~E., Kravtsov, A.~V., Gnedin, N.~Y., Abel, T., \& Rudd, D.~H.
  2009, \mnras, 1774

\bibitem[{Rosenblum {et~al.}(2010)Rosenblum, Garaud, Traxler, \&
  Stellmach}]{Rosenblum:2010p5959}
Rosenblum, E., Garaud, P., Traxler, A., \& Stellmach, S. 2010, arXiv

\bibitem[{Rosswog {et~al.}(2008)Rosswog, Ramirez-Ruiz, \&
  Hix}]{Rosswog:2008p3059}
Rosswog, S., Ramirez-Ruiz, E., \& Hix, W.~R. 2008, \apj, 679, 1385

\bibitem[{Rosswog {et~al.}(2009)Rosswog, Ramirez-Ruiz, \&
  Hix}]{Rosswog:2009p3553}
---. 2009, \apj, 695, 404

\bibitem[{Schlaufman(2010)}]{Schlaufman:2010p4689}
Schlaufman, K.~C. 2010, \apj, 719, 602

\bibitem[{Shen \& Turner(2008)}]{Shen:2008p4684}
Shen, Y. \& Turner, E.~L. 2008, \apj, 685, 553

\bibitem[{Springel(2010)}]{Springel:2010p1909}
Springel, V. 2010, \mnras, 401, 791

\bibitem[{Takeda \& Rasio(2005)}]{Takeda:2005p4492}
Takeda, G. \& Rasio, F.~A. 2005, \apj, 627, 1001

\bibitem[{Tasker {et~al.}(2008)Tasker, Brunino, Mitchell, Michielsen, Hopton,
  Pearce, Bryan, \& Theuns}]{Tasker:2008p4296}
Tasker, E.~J., Brunino, R., Mitchell, N.~L., Michielsen, D., Hopton, S.,
  Pearce, F.~R., Bryan, G.~L., \& Theuns, T. 2008, \mnras, 390, 1267

\bibitem[{Triaud {et~al.}(2010)Triaud, Cameron, Queloz, Anderson, Gillon, Hebb,
  Hellier, Loeillet, Maxted, Mayor, Pepe, Pollacco, S{\'e}gransan, Smalley,
  Udry, West, \& Wheatley}]{Triaud:2010p4772}
Triaud, A. H. M.~J., Cameron, A.~C., Queloz, D., Anderson, D.~R., Gillon, M.,
  Hebb, L., Hellier, C., Loeillet, B., Maxted, P.~F., Mayor, M., Pepe, F.,
  Pollacco, D., S{\'e}gransan, D., Smalley, B., Udry, S., West, R.~G., \&
  Wheatley, P.~J. 2010, arXiv, astro-ph.EP

\bibitem[{Usami \& Fujimoto(1997)}]{Usami:1997p4407}
Usami, M. \& Fujimoto, M. 1997, \apj, 487, 489

\bibitem[{Watson {et~al.}(2010)Watson, Littlefair, Diamond, Cameron,
  Fitzsimmons, Simpson, Moulds, \& Pollacco}]{Watson:2010p4924}
Watson, C., Littlefair, S., Diamond, C., Cameron, A.~C., Fitzsimmons, A.,
  Simpson, E., Moulds, V., \& Pollacco, D. 2010, arXiv

\bibitem[{Zakamska {et~al.}(2010)Zakamska, Pan, \& Ford}]{Zakamska:2010p5800}
Zakamska, N.~L., Pan, M., \& Ford, E.~B. 2010, \mnras, 1566

\bibitem[{Zhou {et~al.}(2007)Zhou, Lin, \& Sun}]{Zhou:2007p5956}
Zhou, J.-L., Lin, D. N.~C., \& Sun, Y.-S. 2007, \apj, 666, 423

\end{thebibliography}

\end{document}